\documentclass[lettersize,journal]{IEEEtran}
\usepackage{amsmath,amsfonts}
\usepackage[table]{xcolor}
\usepackage{algorithmic}
\usepackage{algorithm}
\usepackage{array}
\usepackage{svg}
\usepackage[caption=false,font=normalsize,labelfont=sf,textfont=sf]{subfig}
\usepackage{textcomp}
\usepackage{stfloats}
\usepackage{url}
\usepackage{verbatim}
\usepackage{graphicx}

\usepackage{cite}
\usepackage[printonlyused]{acronym}
\begin{document}

\title{Deep Learning Based Active Spatial Channel Gain Prediction Using a Swarm of Unmanned Aerial Vehicles}

\author{
Enes Krijestorac
and Danijela Cabric,~\IEEEmembership{Fellow,~IEEE}
\thanks{This work was
supported in part by NSF under Grant 1929874.}
\thanks{The authors are with the Electrical and Computer Engineering Department,
University of California at Los Angeles, Los Angeles, CA 90095 USA (e-mail: enesk@ucla.edu;  danijela@ee.ucla.edu).}}



\maketitle

\acrodef{DQN}{deep Q-network}
\acrodef{CG}{channel gain}
\acrodef{MAE}{mean absolute error}
\acrodef{MSE}{mean square error}
\acrodef{RMSE}{root mean square error}
\acrodef{GPR}{Gaussian process regression}
\acrodef{DL}{deep learning}
\acrodef{RL}{reinforcement learning}
\acrodef{UAV}{unmanned aerial vehicle}

\begin{abstract}
Prediction of wireless \ac{CG} across space is a necessary tool for many important wireless network design problems. 
In this paper, we develop prediction methods that use environment-specific features, namely building maps and \ac{CG} measurements, to achieve a high prediction accuracy.
We assume that measurements are collected using a swarm of coordinated unmanned aerial vehicles (UAVs).
We develop novel active prediction approaches which consist of both methods for UAV path planning for optimal measurement collection and methods for prediction of \ac{CG} across space based on the collected measurements. We propose two active prediction approaches based on \ac{DL} and Kriging interpolation. The first approach does not rely on the location of the transmitter and utilizes 3D maps to compensate for the lack of it. We utilize \ac{DL} to incorporate 3D maps into prediction and reinforcement learning for optimal path planning for the UAVs based on \ac{DL} prediction. The second active prediction approach is based on Kriging interpolation, which requires known transmitter location and cannot utilize 3D maps. 
We train and evaluate the two proposed approaches in a ray-tracing-based channel simulator. 
Using simulations, we demonstrate the importance of active prediction compared to prediction based on randomly collected measurements of channel gain. Furthermore, we show that using \ac{DL} and 3D maps, we can achieve high prediction accuracy even without knowing the transmitter location. We also demonstrate the importance of coordinated path planning for active prediction when using multiples UAVs compared to UAVs collecting measurements independently in a greedy manner. 

\end{abstract}

\begin{IEEEkeywords}
channel gain prediction, UAV, deep learning
\end{IEEEkeywords}

\section{Introduction}
\IEEEPARstart{T}{he} use of unmanned aerial vehicles (UAVs) as communication enablers has received a lot of attention in recent years, in part, due to their ability to optimize their placement in order to increase \ac{CG} to the ground devices they are serving \cite{mozaffari2019tutorial}. 
Algorithms for optimal placement often rely on the knowledge of the \ac{CG} across space, which can be obtained via direct measurements or via some type of predictive model.
Prediction of the \ac{CG} across space is also necessary for other important wireless network design problems such as wireless network infrastructure planning \cite{wang2007efficient}, network resource allocation and
spectrum sharing \cite{achtzehn2012improving}.

Commonly used statistical approaches for \ac{CG} prediction rely on the assumption that the channel can be modeled based on features that are not environment-specific, such as distance between radio devices, altitude of radio devices and others. Some examples of statistical UAV communications channel models are provided in \cite{khuwaja2018survey}.
The advantage of using statistical models for spatial channel prediction lies in their computational simplicity and in their suitability for mathematical analysis.
However, such prediction approaches lack the ability to adapt to a given environment which limits the accuracy of their prediction. The key reason for this is that the local environment blockage and scattering may cause the channel to sharply differ from the predictions drawn from simple statistical features such as distance and altitude. 

In order to circumvent the limitations of statistical approaches, methods that utilize environment adaptive features such as 3D maps or \ac{CG} measurements can be utilized. These approaches have the advantage of adapting to the propagation characteristics of the current environment. An example of such methods for predicting the wireless channel is ray-tracing. Ray-tracing can be used to accurately simulate the wireless channel for a specific environment. However, it has the disadvantages of requiring a precise 3D map of the environment, exact transmitter location and is highly computationally complex. An alternative set of methods for environment adaptive spatial prediction are spatial interpolation methods often used in geostatistics, such as Kriging interpolation and inverse distance weighting. These methods can be used to predict the \ac{CG} across an area of interest given a set of sparse measurements \cite{angjelicinoski2011comparative}. 

However, majority of current research on environment adaptive \ac{CG} prediction based on measurements does not consider the methods according to which measurements are collected. For applications such as UAV-enabled communications, it is possible to utilize one or multiple UAVs to collect \ac{CG} measurements for \ac{CG} prediction. For other applications such as network planning, it is also possible to use UAVs or other types of vehicles to collect measurements. The design of paths according to which measurements are collected can significantly influence the accuracy of predicted \ac{CG}. Despite their potential importance, methods for path planning for measurement collection have scarcely been considered in the prior literature. 
Therefore, the first goal of this paper is to develop \ac{CG} prediction approaches that include methods for measurement collection, which we refer to as active \ac{CG} prediction. 

Additionally, spatial interpolation algorithms and ray-tracing \ac{CG} prediction methods rely on exact transmitter location knowledge. Obtaining the exact transmitter location may not always be possible for several reasons:  GPS operation is not always reliable in urban environments, the transmitter equipment may not have localization capabilities, or the transmitter location cannot be shared due to privacy or security reasons. Hence, our second goal is to develop active \ac{CG} prediction approaches that can operate without the knowledge of the exact transmitter location. At the same time, we seek to achieve higher or similar level of accuracy using our location-free \ac{CG} prediction methods compared to traditional spatial interpolation methods, which rely on known transmitter location.

Guided by these objectives, we propose two new active channel gain prediction solutions. Our contributions can be summarized as follows:
\begin{itemize}
    \item  First, we developed an active \ac{DL} \ac{CG} prediction approach that relies on measurements collected by multiple UAVs and a 3D map of the environment, but does not rely on transmitter location. The 3D maps enable highly accurate \ac{CG} prediction compared to spatial interpolation using only measurements and without transmitter location. The developed approach consists of a \ac{DL} prediction method that provides probabilistic \ac{CG} prediction across space and a deep \ac{RL} based method that designs UAV paths for measurement collection for multiple UAVs based on the \ac{DL} predictions.
    This active approach is trained and evaluated in a ray-tracing-based wireless channel simulator. 
    \item  Second, we developed an active \ac{CG} prediction algorithm using multiple UAVs based on Kriging spatial interpolation. 
    While Kriging interpolation has been extensively used for \ac{CG} prediction, no methods for active prediction using multiple UAVs have been proposed. This method is suitable for \ac{CG} prediction when the transmitter location is available and a 3D map of the environment is not. Furthermore, this method does not require extensive training compared to our proposed \ac{DL} active prediction approach. We also evaluated the proposed active Kriging prediction approach in a ray-tracing-based simulator.  
\end{itemize}

The paper is organized as follows. In Sec. \ref{sec:related_work}, we review and compare our work to the existing literature. In Sec. \ref{sec:sys_model}, we provide a detailed description of the targeted application scenarios and the modeling assumptions. In Sec. \ref{sec:active_dl_pred}, we introduce the proposed active \ac{DL} \ac{CG} prediction approach. In sections \ref{sec:deep_learning_predictor} and \ref{sec:deep_rl_controller}, we go into more details on its two main components: probabilistic channel gain prediction and reinforcement learning path planner. In Sec. \ref{sec:active_kriging} we introduce and explain the Kriging based active prediction approach. In Sec. \ref{sec:results}, we describe the simulation environment, benchmarks and obtained simulation results. Finally, in Sec. \ref{sec:conclusions}, we summarize the findings of the paper.

\section{Related Work}
\label{sec:related_work}
The most common approaches for channel prediction are adopted from the field of spatial interpolation \cite{angjelicinoski2011comparative, chowdappa2018distributed, hernandez2012field, braham2016spatial}. Among the interpolation methods, the most commonly used ones are inverse distance weighting (IDW), gradient plus inverse distance squared (GIDS) and Kriging interpolation. Algorithms based on Kriging interpolation rely on the location of the transmitter, while IDW or GIDS based algorithms normally do not, which comes at the cost of lower prediction accuracy compared to Kriging. 
Other stand-alone approaches based on Gaussian modeling of shadowing component of \ac{CG} were proposed in \cite{lee2017channel} and \cite{malmirchegini2012spatial}, but these approaches also rely on knowing the transmitter location. More recently, in \cite{zhang2020spectrum}, thin plate splines (TPS) interpolation method and coupled block-term tensor decomposition methods were used for \ac{CG} prediction. These methods also do not rely on transmitter location.
However, the spatial interpolation based approaches in \cite{angjelicinoski2011comparative, chowdappa2018distributed, hernandez2012field, braham2016spatial} and in \cite{lee2017channel, malmirchegini2012spatial, zhang2020spectrum} do not consider optimal measurement collection methods. Furthermore, these methods are not able to incorporate complex inputs such as topography maps or building maps into their prediction. 

\ac{DL} based approaches have been developed for spatial gain prediction. These approaches are usually developed to outperform spatial interpolation methods in terms of prediction accuracy or to outperform ray tracing approaches in terms of computational complexity, such as in \cite{levie2021radiounet, krijestorac2023agile}, where spatial gain prediction is performed assuming a known 3D map of the environment and location of the transmitter. In \cite{han2020power}, generative adversarial neural networks are used for spatial prediction based on measurements, while the authors in \cite{teganya2020deep} use deep completion auto-encoders for the same task. The approaches in \cite{han2020power} and \cite{teganya2020deep} do not assume to know the user location. 
In our prior work \cite{krijestorac2021spatial}, we also utilized 3D maps and signal strength measurements for probabilistic \ac{CG} prediction using \ac{DL} and this paper builds upon that work.
However, like the approaches in \cite{han2020power, teganya2020deep}, we have not explored optimal \ac{CG} measurement collection strategies.

Optimal measurement collection methods for spatial \ac{CG} prediction have rarely been considered in the prior literature. In \cite{alimpertis2022unified}, the authors consider the problem of cellular base station gain prediction based on crowd-sourced measurements from end-user devices. In this case, the objective is to devise a strategy for optimal selection and utilization of crowd-sourced measurements but not to directly control where the measurements are collected. 
Optimal path planning for measurement collection using UAVs has been considered in \cite{shrestha2022spectrum}. However, in this work, path planning for measurement collection using only a single UAV is considered and machine-learning-based path planning is not considered. Given that the UAV technology is currently mature and widely accessible, it is reasonable to deploy swarms of UAVs for \ac{CG} prediction for one or more transmitters. 
Therefore, we develop path planning methods for multiple coordinated UAVs. Furthermore, it is necessary to consider learning-based methods for path planning for measurement collection due to the recent interest in deep-learning-based \ac{CG} predictors. Since deep neural networks are black box models, it is difficult to design optimal analytical approaches for path planning, hence learning-based approaches can be used to better accomplish this task. 
In \cite{li2023uav}, reinforcement learning was applied for path planning for measurement collection for TPS interpolation. However, this approach was developed for control of a single UAV, so it would not be applicable to coordinated control of multiple UAVs. Furthermore, since it was trained and evaluated for prediction based on TPS interpolation, it is not clear if it would extend to \ac{DL} \ac{CG} prediction approaches.

\section{System model and objectives}

\label{sec:sys_model}

\subsection{Environment and transmitters} We consider a rectangular urban area of interest (AoI) of width $w$, length $l$ and height $h$, for which we have a database of major buildings and objects that can be used to construct a 3D map of the environment, which is denoted by $\mathbf{M}$. 
We discretize the AoI into a uniformly-spaced 3D grid of locations $\mathcal{Q}=\left\{ \mathbf{q}_1,\dots,\mathbf{q}_{|\mathcal{Q}|}\right\}$ with spacing $d$. 
Each location $\mathbf{q}_j \in \mathcal{Q}$ is coordinate vector $\mathbf{q}_j=[q_{j,x},q_{j,y},q_{j,z}]^T$. 
The number of locations in $\mathcal{Q}$ is $\left|\mathcal{Q}\right| = \frac{l}{d} \times \frac{w}{d} \times \frac{h}{d}$.

Throughout this paper, we assume that the goal is to estimate the \ac{CG} in the AoI for a single transmitter $k$.  
The transmitter is assumed to be stationary with a location $\mathbf{w}_{TX}$. 
The location $\mathbf{w}_{TX}$ could be known or unknown, and we will propose approaches that will handle both of these cases.
Furthermore, we assume that \ac{CG} is predicted for a set of points in $\mathcal{Q}_P=\left\{ \tilde{\mathbf{q}}_{1},\dots,\tilde{\mathbf{q}}_{|\mathcal{Q}_P|}\right\} \subset \mathcal{Q}$, which have an altitude $h_{P}$. 
The number of locations in $\mathcal{Q}_P$ is $\left|\mathcal{\mathcal{Q}_P}\right| = \frac{l}{d} \times \frac{w}{d}$.
For the purposes of this paper, the \ac{CG} is predicted for a constant altitude to reduce the training time and computational complexity of the presented \ac{CG} prediction algorithms. However, in applications such as UAV communications, for example, 3D \ac{CG} prediction may be necessary. Therefore, the active \ac{CG} prediction algorithms that will be presented in later sections can naturally be extended to 3D.

The time-averaged narrow-band \ac{CG} in logarithmic scale for a particular transmitter $k$ can be modeled as a function of space $\psi_k(\cdot)$.
The \ac{CG} function $\psi_k(\cdot)$ evaluated across locations $\mathcal{Q}_P$ for user $k$ are stacked into a vector $\mathbf{x}_k \in \mathbb{R}^{|\mathcal{Q}_P|}$. 
We further define a utility binary vector variable $\mathbf{z} \in \mathbb{Z}_2^{|\mathcal{Q}_P|}$, where $\mathbb{Z}_2^R$ denotes the set of all binary integer vectors of size $R$. $[\mathbf{z}]_j=0$ if the location $\tilde{\mathbf{q}}_j \in \mathcal{Q}_P$ is obstructed by a building, i.e. it is indoor, and $[\mathbf{z}]_j=1$ otherwise.

\subsection{UAV swarm and \ac{CG} measurements} We assume that $N$ UAVs are deployed to predict the \ac{CG} in the AoI. We also assume that these UAVs are constrained to move and collect \ac{CG} measurements at a constant altitude $h_{\text{UAV}}$. This assumption is made for the purposes of this paper, to reduce the training time and computational complexity of the presented path planning algorithms. However, in practice, the UAVs would have the ability to move vertically depending on the local flight regulations. Therefore, the path planning algorithms that we will discuss in later sections can be extended to 3D mobility. 

Furthermore, let us denote the placement of UAVs at time $t$, rounded to the nearest grid point in $\mathcal{Q}$, by $\mathbf{P}_t \in \mathcal{Q}^N_{\text{UAV}}$. 
Similarly, we denote the location of each individual UAV, $n$, by $\mathbf{p}_{t,n} \in \mathcal{Q}_{\text{UAV}} \subset \mathcal{Q}$.
We ignore UAV localization errors in our system model. Moreover, we assume that the control of UAVs can be centralized and performed at one of the UAVs or at a nearby edge server. This also implies that UAVs can communicate within the UAV swarm, either directly or through message relaying within the swarm. 

We assume that, as UAVs move, they estimate the \ac{CG} for the transmitter $k$ via the use of pilot signals. 
We assume that time is discretized into time steps, where in each time step, UAVs move by a certain distance and estimate the \ac{CG} at new locations. 
The set of new locations that have been visited by at least one UAV between time $t_1$ and up to and including time $t_2$ is denoted by $\mathcal{V}_{t_1:t_2}=\left\{\mathbf{v}_1, \dots, \mathbf{v}_{|\mathcal{V}_{t_1:t_2}|}
  \right\}$. 
At these locations, the set of measurements of \ac{CG} are obtained via some measurement model $\mathbf{y}_{k,{t_1:t_2}}=C(\mathcal{V}_{t_1:t_2}, \mathbf{x}_k)$. 
Here, we denote the set of \ac{CG} measurements obtained at locations $\mathcal{V}_{t_1:t_2}$ by $\mathbf{y}_{k,{t_1:t_2}}$.
In this paper, we assume a perfect \ac{CG} measurement model, where $\mathbf{y}_{k, {t_1:t_2}}$ measurements are equal to the true channel gains at $\mathcal{V}_{t_1:t_2}$, which we denote by $\mathbf{x}_{k, {t_1:t_2}}$. 

\subsection{Objectives} 
\begin{figure}
    \centering
    \includegraphics[width=0.8\linewidth]{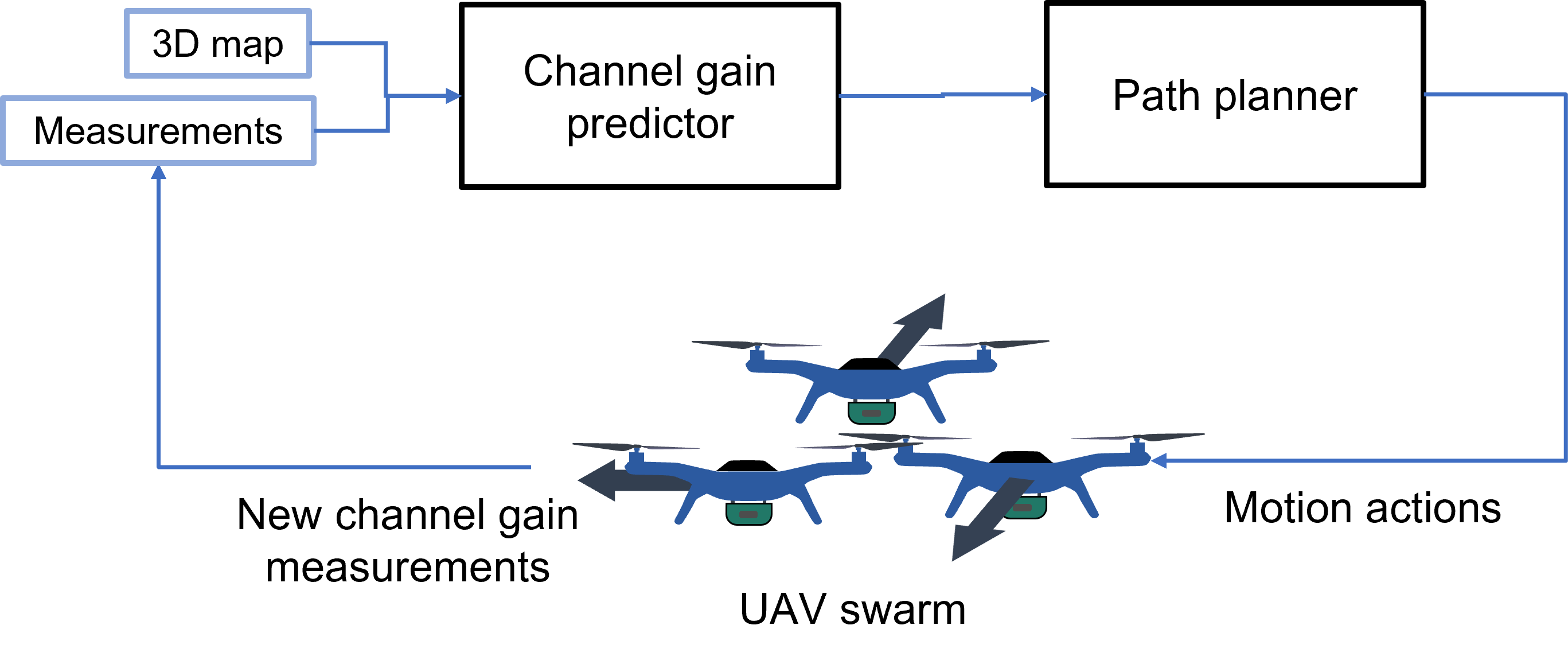}
    \caption{Active \ac{CG} prediction using multiple UAVs}
    \label{fig:active-method}
\end{figure}

The main objective of this paper is to devise active methods for prediction of \ac{CG} $\mathbf{x}_{k}$, as illustrated in Fig. \ref{fig:active-method}.
We seek to develop methods that adapt to the local environment using two kinds of input features: (1) measurements $\mathbf{y}_{k,{1:t}}$ collected by the UAVs up to time time $t$; (2) 3D map $\mathbf{M}$ of the AoI. Given that we seek to develop approaches for prediction of $\mathbf{x}_{k}$ that rely on \ac{CG} measurements, optimally controlling the UAVs to collect measurements that provide maximum amount of information about $\mathbf{x}_{k}$ is just as important as optimally utilizing the measurements to predict $\mathbf{x}_{k}$. Therefore, we seek to develop active \ac{CG} prediction approaches that consist of methods for control of UAVs to collect measurements and methods that can predict $\mathbf{x}_{k}$. 

\section{Deep Learning Based Active \ac{CG} Prediction}
\label{sec:active_dl_pred}

In this section, we explain our \ac{DL} based approach for active \ac{CG} prediction. Our proposed approach consists of two key parts: a deep-learning algorithm to provide a posterior probability prediction of $\mathbf{x}_k$, $p({\mathbf{x}}_k \mid \mathbf{y}_{k,{1:t}}, \mathbf{M})$, and a multi-agent deep reinforcement learning algorithm to control the UAVs to collect measurements based on $p({\mathbf{x}}_k \mid \mathbf{y}_{k,{1:t}}, \mathbf{M})$, 3D map $\mathbf{M}$ and UAV locations.
We propose an iterative procedure where in each time slot, the UAVs move and measure the \ac{CG} at new locations. At the end of each time slot, the measurements from all UAVs are combined to update $p({\mathbf{x}}_k \mid \mathbf{y}_{k,{1:t}}, \mathbf{M})$. 

The posterior $p({\mathbf{x}}_k\mid \mathbf{y}_{k,{1:t}}, \mathbf{M})$ represents the belief on what the true \ac{CG} $\mathbf{x}_k$ is and it is used in two ways in our proposed approach. 
First, the predicted \ac{CG} can be obtained as $\hat{\mathbf{x}}_k = \int_{\mathbf{x}_k} \mathbf{x}_k p(\mathbf{x}_k \mid \mathbf{y}_{k,{1:t}}, \mathbf{M})$. 
Second, the posterior distribution $p(\mathbf{x}_k \mid \mathbf{y}_{k,{1:t}}, \mathbf{M})$ is used to estimate the uncertainty $\text{Var}(\hat{\mathbf{x}}_k)$ of predicted \ac{CG}.
While the posterior probability $p({\mathbf{x}}_k \mid \mathbf{y}_{k,{1:t}}, \mathbf{M})$ can be used to estimate the uncertainty of \ac{CG} prediction across space,  optimally collecting measurements to feed into the \ac{DL} model is still a challenging problem. 
For example, it may not be optimal to simply move UAVs towards locations with high \ac{CG} uncertainty.
This is the case because it is not possible to determine how a deep neural network uses the input measurements to arrive at the final prediction, which is why deep neural networks are often referred to as black box models. 
Furthermore, it is not clear how can multiple UAVs coordinate their movement to optimally collect \ac{CG} measurements.
Therefore, we utilize reinforcement learning to learn optimal path planning algorithms for UAVs that will result in optimal set of \ac{CG} measurements being collected. 

We assume that the UAVs are allotted a fixed number of steps $T$ to estimate $\mathbf{x}_k$. However, an alternative approach is possible where UAVs can perform early stopping of measurement collection and \ac{CG} prediction based on the posterior probability $p({\mathbf{x}}_k \mid \mathbf{y}_{k,{1:t}}, \mathbf{M})$.

The proposed active \ac{DL} \ac{CG} approach is illustrated in Fig. \ref{fig:dl_active_aproach} and its two main components are explained in sections \ref{sec:deep_learning_predictor} and \ref{sec:deep_rl_controller}.
Our proposed \ac{DL} approach for probabilistic \ac{CG} prediction is explained in Sec. \ref{sec:deep_learning_predictor} and our proposed reinforcement learning approach for path planning for measurement collection is explained in Sec. \ref{sec:deep_rl_controller}. 

\section{Deep learning based probabilistic \ac{CG} predictor}
\label{sec:deep_learning_predictor}
The purpose of the proposed \ac{DL} \ac{CG} predictor is to output a posterior distribution  $p(\mathbf{x}_k \mid \mathbf{y}_{k,{1:t}}, \mathbf{M})$ given measurements $\mathbf{y}_{k,{1:t}}$ collected by the UAVs and a 3D map $\mathbf{M}$ of the AoI. 
Deep neural networks are known to be universal function approximators, which is why we use their capabilities to learn the complicated relationship between the 3D map $\mathbf{M}$, \ac{CG} measurements $\mathbf{y}_{k,{1:t}}$ and the \ac{CG} $\mathbf{x}_k$.
To design and train the predictor, we must select a probability distribution that will be used as the posterior $p(\mathbf{x}_k \mid \mathbf{y}_{k,{1:t}}, \mathbf{M})$. 
We selected the Gaussian distribution as the posterior $p(\mathbf{x}_k \mid \mathbf{y}_{k,{1:t}}, \mathbf{M}) = \mathcal{N}(\mathbf{\mu}_k, \mathbf{\Sigma}_k)$ due to its similarity to the Gudmundons model of channel gain shadowing \cite{gudmundson1991correlation}
and because it performed the best in terms of prediction accuracy amongst the alternatives that we tried. 
Two DNN models are used to predict the mean $\mathbf{\mu}_k$ and covariance $\mathbf{\Sigma}_k$, which uniquely define the Gaussian posterior $p(\mathbf{x}_k \mid \mathbf{y}_{k,{1:t}}, \mathbf{M}) = \mathcal{N}(\mathbf{\mu}_k, \mathbf{\Sigma}_k)$.
Furthermore, we restrict the covariance matrix $\mathbf{\Sigma}_k$ to be diagonal to simplify the training loss function, which reduces the computational complexity during training and improves the training convergence rate.

The two deep neural networks that learn the mappings from $(\mathbf{y}_{k,{1:t}}, \mathbf{M})$ to $\mu_k$ and $\mathbf{\Sigma}_k$, 
are denoted as ${\mu}_\theta(\mathbf{y}_{k,{1:t}}, \mathbf{M})$ and $\Sigma_\theta(\mathbf{y}_{k,{1:t}}, \mathbf{M})$, respectively. The parameters of the two DNNs are denoted by $\theta$.


The loss function is based on maximizing the log-likelihood of having observed the training dataset given a Gaussian distribution:
\begin{multline}
    \label{eq:loss_func}
    \mathcal{L}(\theta) = 
    \frac{1}{2D}\sum_{i=1}^D(\Delta_k^{(i)})^T 
    \Sigma_\theta(\mathbf{y}^{(i)}_{k,{1:t}}, \mathbf{M}^{(i)}) ^{-1} 
    \Delta_k^{(i)}   + \\
     \frac{1}{2D} \sum_{i=1}^D \log \left( (\mathbf{z}_k^{(i)})^T \text{diag}\left( \Sigma_\theta(\mathbf{y}^{(i)}_{k,{1:t}}, \mathbf{M}^{(i)}) \right) \right)
\end{multline}
where $D$ is the number of training samples. In Eq. \ref{eq:loss_func}, we used a substitute variable $\Delta_k^{(i)} = \left({\mu}_\theta(\mathbf{y}^{(i)}_{k,{1:t}}, \mathbf{M}) -\mathbf{x}_k^{(i)} \right)\odot \mathbf{z}_k^{(i)}$. The loss function considers only outdoor coordinates through the use of $\mathbf{z}_k$ in the expression. The variable $\mathbf{z}_k$ must be known for training but not for prediction during deployment.
We minimize the loss function in Eq. \ref{eq:loss_func} to optimize the parameters of the deep neural networks: $\theta = \arg\min_{\theta}\mathcal{L}(\theta)$. 

\subsection{Deep neural network design} 
\begin{figure*}
    \centering
    \includegraphics[width=0.8\linewidth]{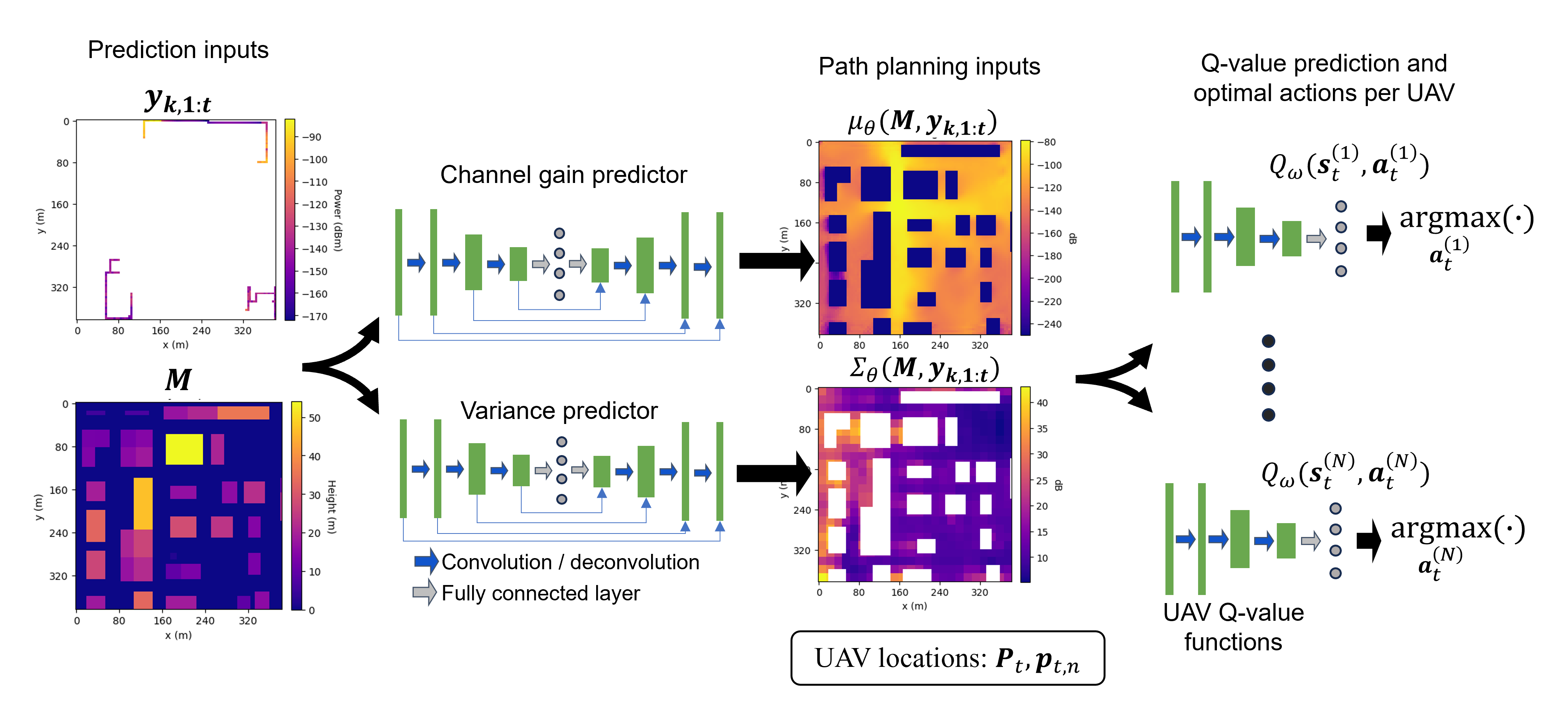}
    \caption{The proposed active \ac{DL} \ac{CG} prediction approach consists of two main components: \ac{DL} probabilistic predictor and a \ac{RL} path planner. The predictor outputs mean and variance of the posterior distribution of the \ac{CG}, ${\mu}_\theta(\mathbf{M},\mathbf{y}_{k,{1:t}})$ and ${\Sigma}_\theta(\mathbf{M},\mathbf{y}_{k,{1:t}})$, given the channel gain measurements $\mathbf{y}_{k,{1:t}}$ and a 3D map $\mathbf{M}$. UAV path planning is based on on reinforcement learning. Each UAV uses a \ac{DQN} to predict the value of an action in a particular state and then select the optimal action accordingly. The \ac{DQN} inputs are locations of the UAVs, ${\mu}_\theta(\mathbf{M},\mathbf{y}_{k,{1:t}})$ and ${\Sigma}_\theta(\mathbf{M},\mathbf{y}_{k,{1:t}})$.}
    \label{fig:dl_active_aproach}
\end{figure*}

We use a convolutional neural network architecture for deep neural networks ${\mu}_\theta(\mathbf{y}_{k,{1:t}}, \mathbf{M})$ and $\Sigma_\theta(\mathbf{y}_{k,{1:t}}, \mathbf{M})$. 
Convolutional neural networks are suitable for task of spatial \ac{CG} prediction because convolutional neural network layers consist of spatial filters that enforce a local connectivity pattern between neurons of adjacent layers. This architecture ensures that the learned filters produce the strongest response to spatially local \ac{CG} measurements. 
The particular convolutional neural network architecture that we used was U-Net as illustrated in Fig. \ref{fig:dl_active_aproach}, which was first used for image segmentation problems \cite{ronneberger2015u}.
U-Net architecture is particularly suitable for \ac{CG} prediction due to skip connections, shown in Fig. \ref{fig:dl_active_aproach}, which enable 3D map and \ac{CG} measurements to be passed to the final layers without information loss due to encoding. 
To use convolutional neural networks for this problem, we convert the inputs $(\mathbf{y}_{k,{1:t}}, \mathbf{M})$ into matrices that can be processed by convolutional neural networks, as shown in Fig. \ref{fig:dl_active_aproach}. First, the 3D map $\mathbf{M}$ is converted into a tensor $\tilde{\mathbf{M}} \in \mathbb{R}^{\frac{l}{d} \times \frac{w}{d}}$, where each entry $\left[\tilde{\mathbf{M}}\right]_{i,j}$ is equal to the building or terrain height at coordinates $(id, jd)$. 
Next, the measurements collected up to time $t$ are converted into a matrix $\tilde{\mathbf{Y}}_{k,1:t} \in \mathbb{R}^{\frac{l}{d} \times \frac{w}{d}}$ , where each entry $\left[\tilde{\mathbf{Y}}_{k,1:t}\right]_{i,j}$ is equal to the \ac{CG} measurement at coordinates $(id, jd, h_{\text{UAV}})$, if $(id, jd, h_{\text{UAV}}) \in \mathcal{V}_{1:t}$, and is equal to $c_L$, otherwise. $c_L$ is a padding value that is set to a value outside of the reasonable range of \ac{CG} values. The outputs of ${\mu}_\theta(\mathbf{y}_{k,{1:t}}, \mathbf{M})$ and $\Sigma_\theta(\mathbf{y}_{k,{1:t}}, \mathbf{M})$ are also matrices of form $\mathbb{R}^{\frac{l}{d} \times \frac{w}{d}}$, which are transformed into a vector $\mathbb{R}^{|\mathcal{Q}_P|}$ and a diagonal matrix $\mathbb{R}^{|\mathcal{Q}_P| \times |\mathcal{Q}_P|}$, respectively.  

\section{Deep reinforcement learning for optimal measurement collection}
\label{sec:deep_rl_controller}
In this section, we describe our \ac{RL} approach for control of UAVs to optimally collect \ac{CG} measurements, which is part of our active \ac{CG} prediction framework, as illustrated in Fig. \ref{fig:dl_active_aproach}.

\subsection{Path planning problem formulation} 
\label{sec:path-planning-problem-formulation}
We formulate the trajectory design problem as a sequential decision making problem. At each time step $t$, the path planning controller will define an action $\mathbf{u}_{t} $. 
The action $\mathbf{u}_t$ specifies the next displacement for each of the $N$ UAVs. We limit the number of possible displacements to $U=4$, where each displacement is of length $d$ and along one of the horizontal directions. 
Therefore, the motion actions are discrete and number of feasible motion actions across the entire UAV swarm is $U^N$.
We define the rules according to which UAVs move using a transition function $\mathbf{P}_{t+1}=T(\mathbf{P}_{t}, \mathbf{u}_{t})$, which defines the positions of UAVs at time $t+1$, $\mathbf{P}_{t+1}$, given the positions of UAVs at time $t$, $\mathbf{P}_{t}$, and motion action $\mathbf{u}_t$.
The action at time step $t$ is constrained to $\mathbf{u}_t \in G(\mathbf{P}_t)$ due to buildings and other obstacles. The function $G(\mathbf{P}_t)$ can be used to capture other constraints on motion of UAVs such as no-fly zones.

Given the above notation, we can define the trajectory design as an optimization problem:
\begin{subequations}
\begin{align} 
\label{eq:P1}
\tag{P1}
\max_{\mathbf{u}_{1}, \dots, \mathbf{u}_{T}} & \quad   
\frac{1}{|\mathcal{Q_P}|}  \left| \left|{\mu}_\theta(\mathbf{y}_{k,{1:T+1}}, \mathbf{M}) -\mathbf{x}_k \right| \right|_2^2 \\
\notag
\textrm{s.t.} & \quad 
\mathbf{P}_{t+1}=T(\mathbf{P}_{t}, \mathbf{u}_{t}), \mathbf{u}_t \in G(\mathbf{P}_t)
\end{align}
\end{subequations}
where the objective function is the \ac{MSE} of the \ac{CG} prediction after $T$ time steps. This is a challenging problem to solve because we cannot predict how the motion actions $\mathbf{u}_{1}, \dots, \mathbf{u}_{T}$ will influence the \ac{MSE}. Therefore, we turn to reinforcement learning to learn the relationship between $\mathbf{u}_{1 }, \dots, \mathbf{u}_{T}$ and the \ac{MSE} $ \frac{1}{|\mathcal{Q_P}|} \left| \left|{\mu}_\theta(\mathbf{y}_{k,{1:T+1}}, \mathbf{M}) -\mathbf{x}_k \right| \right|_2^2$, to solve for optimal actions $\mathbf{u}_{1}, \dots, \mathbf{u}_{T}$.

\subsection{Reinforcement learning background} 
\newcommand{\Step}{t}
\newcommand{\MState}{\mathbf{s}}
\newcommand{\MStateSpace}{\mathcal{S}}
\newcommand{\Action}{\mathbf{a}}
\newcommand{\ActionSpace}{\mathcal{A}}
\newcommand{\Environment}{\mathcal{E}}

Reinforcement learning is a branch of machine learning that is concerned with making sequences of decisions. It can be applied to problems that can be casted as a Markov decision process (MDP). 
A Markov decision process (MDP) is defined by a tuple $\{\MStateSpace, \ActionSpace, {P}, r\}$, where $\MStateSpace$ is a set of states, $\ActionSpace$ is a set of actions, ${P}$ is the transition probability function from state $\MState$ to $\MState^{\prime} \in \MStateSpace$ after action $\Action \in \ActionSpace$ is performed, $r$ is the reward obtained after $\Action$ is executed in state $\MState$. An action space $\ActionSpace$ can be a discrete or a continuous set.  

A policy $\pi: \mathcal{S} \rightarrow \mathcal{A}$ is a function that maps a state $\MState \in \mathcal{S}$ into an action $\Action \in \mathcal{A}$. 
With some abuse of notation, we also use variable $t$ to denote the time-step in the MDP.
Thus, at time $t$, the agent observes the state $\MState_t$, then based on a specific policy $\pi$, it takes action $\Action_t=\pi\left(\MState_\Step\right)$. 
Consequently, a new state $\MState_{\Step+1}$ will be reached with probability ${P}\left(\MState_{\Step+1} \mid \MState_t, \Action_t\right)$ and a reward $r_{t}$ will be received. 
The observed information from the environment, the reward $r_{t}$ and $\MState_{\Step+1}$ are used to improve the policy. 
This process is repeated until the optimal policy is reached. 
We use $\rho_\pi\left(\mathbf{s}_\Step\right)$ and $\rho_\pi\left(\MState_\Step, \mathbf{a}_\Step\right)$ to denote the state and state-action probability distributions induced by a policy $\pi$.

The objective function in reinforcement learning is normally the expected sum of rewards 
$\sum_\Step \mathbb{E}_{\left(\MState_\Step, \Action_\Step \right) \sim \rho_\pi}\left[r\left(\MState_\Step, \Action_\Step\right)\right]$ , where $\gamma$ is the discount factor. 
Reinforcement learning methods can broadly be classified into two categories: policy learning and Q-learning. 
In policy learning methods, the goal is to directly learn the optimal policy function $\pi$. 
In Q-learning methods, the goal is to learn the Q-value function 
$
{Q}(\MState, \Action)=\mathbb{E}_{\left(\MState_\Step, \Action_\Step \right) \sim \rho_\pi} \left[\sum_{t=0}^{\infty} \gamma^t r\left(\MState_\Step, \Action_\Step\right) \mid \MState_0=\MState, \Action_0=\Action \right]
$
, which defines the expected reward in a state $\MState$, after taking the action $\Action$.
Based on the Q-value function, the optimal policy is then $\pi(\MState) = {\text{argmax}_a}~{Q}(\MState, \Action)$. 
Deep Q-learning is an extension to the Q-learning paradigm whereby a deep neural network is used to approximate $Q(\MState, \Action)$.

\subsection{Path planning as a Markov decision process} 
Next, we convert the path planning problem described in Sec. \ref{sec:path-planning-problem-formulation} into an MDP. Based on the problem formulation (\ref{eq:P1}), if we were to define the MDP such that the entire UAV swarm is considered a single agent, the size of the action space would grow exponentially with the number of UAVs. While such MDP formulation would be suitable for solving the path planning problem (\ref{eq:P1}), for a large number of UAVs, the size of the actions space would prevent effective training of reinforcement learning policies. Therefore, we focus on decentralized control where each UAV will act as an independent agent, while treating the rest of the UAVs as part of the environment. 
Even though their control is decentralized in our approach, the UAVs within the swarm are still cooperative and share the same common goal stated in the objective function in (\ref{eq:P1}).
Therefore, our problem can be formulated as a multi-agent reinforcement learning problem and the respective MDP formulation is explained next. 

\subsubsection{State space}
Since the state of the environment is not fully observable, a set of observations replaces the role of the state in our MDP formulation. 

In multi-agent reinforcement learning, the input observations often consist of observation related to the environment and messages emitted by other agents that assist the agents in collaboratively achieving the common goal.  
One of the main challenges related to multi-agent reinforcement learning is designing or learning communication protocols between the agents \cite{foerster2018deep}. 
In our approach, we do not aim to learn the messages to be passed between the agents but instead utilize the deep-learning \ac{CG} predictor to enable cooperation between the UAVs. 
The \ac{DL} predictor processes the information collected by the UAVs, namely the \ac{CG} measurements collected by the UAVs and their locations, and outputs features ${\mu}_\theta(\mathbf{y}_{k,{1:t}}, \mathbf{M})$ and ${\Sigma}_\theta(\mathbf{y}_{k,{1:t}}, \mathbf{M})$. The features ${\mu}_\theta(\mathbf{y}_{k,{1:t}}, \mathbf{M})$ and ${\Sigma}_\theta(\mathbf{y}_{k,{1:t}}, \mathbf{M})$ are part of the observation $\mathbf{s}_t^{(n)}$ observed by UAV $n$ at time $t$. Additionally, each UAV $n$ observes its own current location, location of other UAVs in the swarm, and a 3D map of the environment $\mathbf{M}$. In summary, the observation for UAV $n$ at time $t$ is $\MState_\Step^{(n)} = \left({\mu}_\theta(\mathbf{y}_{k,{1:t}}, \mathbf{M}),  {\Sigma}_\theta(\mathbf{y}_{k,{1:t}}, \mathbf{M}), \mathbf{M}, \mathbf{P}_t, \mathbf{p}_{t,n}\right)$. 

\subsubsection{Action space} Since the control of the UAVs is distributed, the action $\Action_t^{(n)}$ dictates the motion of the UAV $n$ at time $t$. Each action maps to one of the $U$ the possible displacements defined in Sec. \ref{sec:path-planning-problem-formulation}. 

\subsubsection{Reward} The reward function is designed to maximize the objective function in (\ref{eq:P1}) and the agents receive the reward $r_t^{(n)}$ at time $t$. The agent receives a reward $r^{(n)}(t) = r_e e^{(n)}(t)$ when $1 \leq t < T$. $e^{(n)}(t)$ is equal to $1$ if the UAV $n$ visits a new location at time $t$ and is 0 otherwise. The exploration reward is useful during training to ensure that UAVs don't visit the same location multiple times. At $t=T$,  the reward is based on the prediction error, $r_t^{(n)}=- r_r \left| \left|{\mu}_\theta(\mathbf{y}_{k,{1:T+1}}, \mathbf{M}) -\mathbf{x}_k \right| \right|_2$. The constants $r_e$ and $r_r$ were empirically selected during training.

\subsection{Deep Q-learning algorithm}

The deep Q-learning algorithm that we developed to solve the MDP described in the previous section is based on the \ac{DQN} algorithm \cite{mnih2013playing}. 
In \ac{DQN}, the Q-value function is approximated by a neural network $Q_\omega$, with parameters $\omega$. 
An estimate of the true Q-value at time $t$, $Q_t$, can be obtained by using a single sample estimate of the Bellman backup operator
\begin{equation}
    \label{eq:bellman}
    \widehat{\mathcal{T}Q_t} = r_t + \underset{\Action_{t+1}}{\text{max}}~\gamma Q_\omega (\MState_{t+1},\Action_{t+1})
\end{equation}
This is called a single sample estimate because only the reward $r_t$ at the current time instant $t$ is used to approximate the infinite horizon Q-value function. 

In order to train $Q_\omega$ to approximate $Q$, the following minimization is done over sample data,
\begin{equation}
    \underset{\omega}{\text{minimize}} \sum_t \left|\left| \widehat{\mathcal{T}Q_t} - Q_\omega(\MState_t, \Action_t)\right|\right|^2
    \label{eq:learn}
\end{equation}
In the DQN algorithm, the training and the interaction of the agent with the environment happen in parallel. 
As the agent gathers experience, samples of that experience are stored and the minimization in the Eq. \ref{eq:learn} is done periodically, every $\tau_L$ steps, by randomly sampling a batch of $B_L$ recorded samples and applying gradient descent. 
This is referred to as experience replay. Each sample is a tuple $(\MState_t, \Action_t, r_t, \MState_{t+1})$ and these are stored in the replay buffer. 

The agent interacts with the environment following the $\epsilon$-greedy policy, where at any time $t$ the agent either takes a random action at probability $\epsilon$ or the Q-value optimal action $\text{argmax}_{\mathbf{a}_t}~Q_\omega(\MState_t,\Action_t)$ at probability $(1-\epsilon)$. 
In the implementation of DQN, there is an additional $Q_\omega$, called the target Q-network. 
The target Q-network is used in the Bellman backup operator but it is not directly optimized over. Instead, its parameters are copied from the main Q-network at period $\tau_{DQN}$. The target Q-network is included to improve the stability during training.

Since its original inception, several modifications of the original \ac{DQN} algorithm have been shown to improve performance over a variety of tasks. In this paper, we apply two of such modifications we found to be useful on our problem: multi-step learning and distributional \ac{RL} \cite{bellemare2017distributional}. In multi-step learning, Bellman backup operator in Eq. \ref{eq:bellman} is extended to include reward samples from $M$ consecutive steps:
\begin{equation}
    \widehat{\mathcal{T}Q_t} = \sum_{m=0}^{M-1} \gamma^m r_{t+m} + \underset{\Action_{t+M}}{\text{max}}~\gamma^M Q_\omega (\MState_{t+M},\Action_{t+M})    
\end{equation}
In distributional RL, the \ac{DQN} is trained to predict a discrete distribution of Q-values on a discrete support $\mathbf{v}$, where $\mathbf{v}$ is vector with $N_a$ atoms. To accomplish this, the DQN is modified to have $N_a \times U$ outputs. Furthermore, the loss functions in Eq. \ref{eq:learn} is replaced by a loss function that ensures that the predicted distribution closely matches the actual distribution of returns, the details of which are omitted for brevity.

\subsection{Multi-agent deep Q-learning for measurement collection using multiple UAVs}

We extended the DQN algorithm to multiple agents in the following way. First, all of the agents or UAVs share the same \ac{DQN} network $Q_\omega$ with identical parameters $\omega$, but each agent acts differently due to different input observations. The policy is trained in a centralized way, where the training samples from all agents are collected and fed to a centralized replay buffer, which is used to train the common policy $Q_\omega$. Similarly to the single-agent DQN, each agent retains a copy of the main \ac{DQN} $Q_\omega$ to collect training samples and these copies are periodically synced with main \ac{DQN} $Q_\omega$. Overall, the changes we had made to the single-agent DQN algorithm are minimal because we use the deep-learning \ac{CG} predictor to process information collected by the UAVs and extract features which are relevant for maximizing the reward obtained. 
These shared input features facilitate collaboration.
Furthermore, since all of the agent Q-networks share the same parameters there is an indirect knowledge transfer
in the parameter space between the agents, which also enables collaboration.

\subsection{Deep Q-network design}

We used a combination of convolutional and fully-connected layers for the design of our \ac{DQN}. 
Convolutional neural networks are suitable for this task because the outputs ${\mu}_\theta(\mathbf{y}_{k,{1:t}}, \mathbf{M})$ and ${\Sigma}_\theta(\mathbf{y}_{k,{1:t}}, \mathbf{M})$ provided by deep learning predictors are matrices as shown in Fig. \ref{fig:dl_active_aproach}. Similarly, the location information $\mathbf{P}_t$ and $\mathbf{p}_{t,n}$ can be converted into binary matrices, with non-zero entries corresponding to locations of the UAVs. 
Using convolutional layers, we can efficiently extract lower dimensional features from the inputs, which are utilized by fully-connected layers for Q-value prediction. The last layer has $N_a \times U$ outputs, which correspond to distributional prediction of Q-value.

\section{Active channel prediction based on Kriging interpolation}
\label{sec:active_kriging}

Kriging interpolation has been extensively used for \ac{CG} prediction. However, no methods for active prediction approaches based on Kriging interpolation using multiple UAVs have been proposed in prior literature. While our \ac{DL} active prediction approach can accomplish \ac{CG} prediction without transmitter location, traditional interpolation methods such as Kriging remain useful when transmitter location is available. Moreover, the Kriging method has the advantage of not requiring extensive training compared to \ac{DL} approaches and is therefore easier to deploy. Hence, in this section, we develop an active \ac{CG} prediction algorithm using multiple UAVs based on Kriging spatial interpolation. 

\subsection{Kriging interpolation for channel gain prediction}
\label{sec:kriging_interpolation}
Kriging interpolation is an equivalent method to \ac{GPR}, which is a widely used method for interpolation, classification, supervised
learning, and active learning \cite{rasmussen2003gaussian}. 
\ac{GPR} constructs a probabilistic prediction of a partially observed function (of time and/or space) assuming this function is a realization of a Gaussian process (GP). 

In statistical models of the \ac{CG}, time-averaged \ac{CG} $\psi_k(\mathbf{q}_j)$ at location $\mathbf{q}_j$ is split into two components, $\psi_k(\mathbf{q}_j) = \psi_{k,PL}(\mathbf{q}_j)+\psi_{k,SH}(\mathbf{q}_j)$, where $\psi_{k, PL}$ is the path loss due to free space attenuation and $\psi_{k, SH}(\mathbf{q}_j)$ is the loss due to shadowing. $\psi_{k, PL}(\mathbf{q}_j)$ can be predicted knowing the antenna radiation pattern and separation of the receiver and the transmitter. On the other hand, $\psi_{k, SH}(\mathbf{q}_j)$ is often modeled as a Gaussian random variable with exponentially decaying spatial correlation according to the Gudmundson model \cite{gudmundson1991correlation}. Accordingly, Kriging interpolation or \ac{GPR} can be used to predict $\psi_{k, SH}(\mathbf{q}_j)$, while the $\psi_{k, PL}(\mathbf{q}_j)$ component can be obtained knowing the distance of location $\mathbf{q}_j$ to the transmitter.

Using Kriging interpolation, we aim to predict the shadowing gain $\tilde{\mathbf{x}}_{k, 1:t}$ at unvisited locations $\mathcal{Q}_P \backslash \mathcal{V}_{1:t}$. 
Then, we can obtain the \ac{CG} at locations $\mathcal{Q}_P \backslash \mathcal{V}_{1:t}$ by adding $\tilde{\mathbf{x}}_{k, 1:t}$ to the estimated free-space path loss gain.
In simple Kriging, the data is modeled as a GP with a zero mean and a
prescribed form of the stationary covariance function (also known as kernel). This modelling is compatible with the Gudmundson shadowing model, therefore we will utilize simple Kriging as the foundation of our active prediction approach.
The path-loss $\psi_{k,PL}(\mathbf{q}_j)$ is estimated using the model:
\begin{equation}
    \label{eq:pl}
    \psi_{k,PL}(\mathbf{q}_j) = \alpha   - \beta \log \left( \left|\left| \mathbf{q}_j-\mathbf{w}_{TX}\right|\right|_2  \right)
\end{equation}
where the constants $\alpha$ and $\beta$ are estimated from \ac{CG} data by minimizing the mean square error loss.

The kernel defines the shadowing gain cross-covariance between two locations $\mathbf{q}_i$ and $\mathbf{q}_j$:
$
    k(\mathbf{q}_i,\mathbf{q}_j) = \text{Cov}\left( \psi_{k,SH}(\mathbf{q}_i), \psi_{k,SH}(\mathbf{q}_j)\right).
$
The kernel we used is based on the Gudmundson model:
\begin{equation}
    \label{eq:kernel}
    k(\mathbf{q}_i,\mathbf{q}_j) = \phi \exp \left( \frac{-\left|\left| \mathbf{q}_i-\mathbf{q}_j\right|\right|_2}{\delta} \right),
\end{equation}
where $\phi$ and $\delta$ are positive constants that are estimated from \ac{CG} data via negative log-likelihood minimization.
The kernel $k(\mathbf{q}_i,\mathbf{q}_j)$ is isotropic, i.e. cross-covariance only depends on distance between $\mathbf{q}_i$ and $\mathbf{q}_j$, but not on the specific values of $\mathbf{q}_i$ and $\mathbf{q}_j$. Given that UAVs have visited a set of locations $\mathcal{V}_{1:t}$ up to time $t$, a vector $\tilde{\mathbf{y}}_{k,1:t}$ of shadowing gain measurements will be obtained. The shadowing gain measurements $\tilde{\mathbf{y}}_{k, 1:t}$ are obtained by subtracting the estimated free-space path-loss gain obtained using Eq. \ref{eq:pl} from ${\mathbf{y}}_{k, 1:t}$.


The covariance matrix of the observed shadowing gains $\tilde{\mathbf{y}}_{k,1:t}$ is denoted by $\mathbf{\Sigma}_{v,v} = \text{Cov}\left(\tilde{\mathbf{y}}_{k,1:t}, \tilde{\mathbf{y}}_{k,1:t} \right)$. The matrix $\tilde{\mathbf{y}}_{k,1:t}$ can be obtained using the kernel as follows:
\begin{equation}
\mathbf{\Sigma}_{v,v}=\left[\begin{array}{ccc}
k(\mathbf{v}_{1},\mathbf{v}_{1}) & \cdots & k(\mathbf{v}_{1},\mathbf{v}_{|\mathcal{V}_{1:t}|})\\
\vdots & \ddots & \vdots\\
k(\mathbf{v}_{|\mathcal{V}_{1:t}|},\mathbf{v}_{1}) & \cdots & k(\mathbf{v}_{|\mathcal{V}_{1:t}|},\mathbf{v}_{|\mathcal{V}_{1:t}|})
\end{array}\right].
\label{eq:sigma_vv}
\end{equation}
Furthermore, we introduce a matrix $\mathbf{\Sigma}_{v,p}$, which denotes the cross-covariance of the shadowing gains between measured locations and prediction locations: $\mathbf{\Sigma}_{v,p} = \text{Cov}\left(\tilde{\mathbf{y}}_{k,1:t}, \tilde{\mathbf{x}}_{k, 1:t} \right)$. The cross-covariance matrix $\mathbf{\Sigma}_{v,p}$ can be obtained using the shadowing kernel in Eq. \ref{eq:kernel}. Finally, we define the covariance matrix $\mathbf{\Sigma}_{p,p}$ of shadowing gain at predicted locations: $\mathbf{\Sigma}_{p,p} = \text{Cov}\left(\tilde{\mathbf{x}}_{k}, \tilde{\mathbf{x}}_{k} \right)$.

Using the matrices $\mathbf{\Sigma}_{v,v}$ and $\mathbf{\Sigma}_{v,p}$, and assuming that the shadowing gain is a zero-mean GP, we can predict $\tilde{\mathbf{x}}_{k, 1:t}$ given $\tilde{\mathbf{y}}_{k,1:t}$ as follows:
\begin{equation}
    \tilde{\mathbf{\mu}}_{k, 1:t} = \mathbf{\Sigma}_{v,p}^T \mathbf{\Sigma}_{v,v}^{-1} \tilde{\mathbf{y}}_{k,1:t}
\end{equation}
Similarly, we can calculate the conditional covariance of the predicted shadowing gains given $\tilde{\mathbf{y}}_{k,1:t}$ as:
\begin{equation}
   \tilde{\mathbf{\Sigma}}_{k, 1:t}= \mathbf{\Sigma}_{p,p} -  \mathbf{\Sigma}_{v,p}^T \mathbf{\Sigma}_{v,v}^{-1} \mathbf{\Sigma}_{v, p}
   \label{eq:kriging_variance}
\end{equation}
The predicted covariance $\tilde{\mathbf{\Sigma}}_{k, 1:t}$ depends on $\mathbf{\Sigma}_{v,v}$, $\mathbf{\Sigma}_{v,p}$ and $\mathbf{\Sigma}_{p,p}$, which are only dependent on the visited locations $\mathcal{V}_{1:t}$ and unvisited locations $\mathcal{Q}_P \backslash \mathcal{V}_{1:t}$ (see for example the definition of $\mathbf{\Sigma}_{v,v}$ in Eq. \ref{eq:sigma_vv}). Therefore, $\tilde{\mathbf{\Sigma}}_{k, 1:t}$ only depends on the selection of the explored locations and not the observed measurements $\tilde{\mathbf{y}}_{k,1:t}$.

\subsection{Optimal path planning for measurement collection}

Next, we develop optimal path planning methods for \ac{CG} measurement collection. The goal of optimal path planning remains to minimize the \ac{MSE} $\frac{1}{|\mathcal{Q_P}|} ||\tilde{\mathbf{\mu}}_{k, 1:T+1}-\tilde{\mathbf{x}}_{k, 1:T+1}||_2^2$. 
This problem could also be solved using \ac{RL}. However, since Kriging interpolation  unlike \ac{DL} prediction is not a data-driven method, we aim to develop optimal path planning methods which also do not rely on large data for this approach.
The \ac{MSE} is not useful for UAV path design since it is unknown to the UAVs. 
Instead, other criteria that are found to strongly correlate to minimizing the mean square error are utilized for sensing of GPs, such as the entropy of $\tilde{\mathbf{y}}_{k, 1:T+1}$ \cite{krause2008near}. 
Since $\tilde{\mathbf{y}}_{k, 1:T+1}$ has a Gaussian distribution, this entropy can be calculated as:
\begin{equation}
H(\tilde{\mathbf{y}}_{k, 1:T+1}) =\frac{| \mathcal{V}_{1:T+1}|}{2} \log(2\pi e) + \frac{1}{2}\log |\mathbf{\Sigma}_{v,v}|    
\label{eq:entropy_metric}
\end{equation}
The purpose of the metric in Eq. \ref{eq:entropy_metric} can be explained as follows.
Using the chain rule for entropy, $H(\tilde{\mathbf{x}}_{k, 1:T+1} , \tilde{\mathbf{y}}_{k, 1:T+1}) = H(\tilde{\mathbf{x}}_{k, 1:T+1}  \mid \tilde{\mathbf{y}}_{k, 1:T+1}) + H(\tilde{\mathbf{y}}_{k, 1:T+1})$.
Since $H(\tilde{\mathbf{x}}_{k, 1:T+1} , \tilde{\mathbf{y}}_{k, 1:T+1})$ is a constant as a function of measured locations, by maximizing $H(\tilde{\mathbf{y}}_{k, 1:T+1})$, the conditional entropy $H(\tilde{\mathbf{x}}_{k, 1:T+1}  \mid \tilde{\mathbf{y}}_{k, 1:T+1})$ is minimized and so is $|\tilde{\mathbf{\Sigma}}_{k, 1:t}|$.

Based on the entropy metric in Eq. \ref{eq:entropy_metric}, we can formulate the path planning problem as: 
\begin{subequations}
\begin{align} 
\label{eq:P2}
\tag{P2}
\max_{\mathbf{u}_{1}, \dots, \mathbf{u}_{T}} & \quad    H(\tilde{\mathbf{y}}_{k, 1:T+1})~
\textrm{s.t.} ~
\mathbf{P}_{t+1}=T(\mathbf{P}_{t}, \mathbf{u}_{t}), \mathbf{u}_t \in G(\mathbf{P}_t)
\end{align}
\end{subequations}
The problem (\ref{eq:P2}) is known to be NP-hard and can only be optimally solved using an exhaustive algorithm.  However, the number of possible paths exponentially increases with $T$ and $N$, so an exhaustive approach becomes intractable for real-time applications. Therefore, it is necessary to develop a suboptimal tracktable heuristic. 

We utilize the derivation in \cite{low2011active} to recast the problem (\ref{eq:P2}) into a deterministic MDP.
Let us denote the measurements collected by the UAVs at time $i$ as $\tilde{\mathbf{y}}_{k, i}$. Then, using the chain rule for entropy, we can rewrite the entropy of the shadowing gains at measured locations as:
$
H(\tilde{\mathbf{y}}_{k, 1:T+1}) = H(\tilde{\mathbf{y}}_{k, 1})+\sum_{t=1}^{T} H(\tilde{\mathbf{y}}_{k, t+1} \mid \tilde{\mathbf{y}}_{k, 1:t})
$. By approximating $H(\tilde{\mathbf{y}}_{k, t+1} \mid \tilde{\mathbf{y}}_{k, 1:t})$ by an upper bound  $H(\tilde{\mathbf{y}}_{k, t+1} \mid \tilde{\mathbf{y}}_{k, t})$, we can simplify our objective function to be
\begin{equation}
\label{eq:entropy_lower_bound}
H(\tilde{\mathbf{y}}_{k, 1:T+1}) \approx H(\tilde{\mathbf{y}}_{k, 1})+\sum_{t=1}^{T} H(\tilde{\mathbf{y}}_{k, t+1} \mid \tilde{\mathbf{y}}_{k, t})
\end{equation}
Then, the new optimization problem based on the approximation in Eq. \ref{eq:entropy_lower_bound} is:
\begin{subequations}
\begin{align} 
\label{eq:P3}
\tag{P3}
\max_{\mathbf{u}_{1}, \dots, \mathbf{u}_{T}} & \quad    \sum_{t=1}^{T} H(\tilde{\mathbf{y}}_{k, t+1} \mid \tilde{\mathbf{y}}_{k, t}) \\
\notag
\textrm{s.t.} & \quad 
\mathbf{P}_{t+1}=T(\mathbf{P}_{t}, \mathbf{u}_{t}), \mathbf{u}_t \in G(\mathbf{P}_t)
\end{align}
\end{subequations}
where we have omitted the term $H(\tilde{\mathbf{y}}_{k, 1})$ from the objective function since it does not depend on $\mathbf{u}_{1}, \dots, \mathbf{u}_{T}$. This objective function leads to paths with actions such that entropy of locations explored at time $t+1$ given the shadowing gain measurements at time $t$ is maximized. 

The problem (\ref{eq:P3}) can be converted into a deterministic MDP, where the state at time $t$ is simply $\mathbf{s}_{t}=\mathbf{P}_{t}$ and the action is $\mathbf{a}_{t}=\mathbf{u}_{t}$.
The reward function at time $t$ is defined as:
\begin{equation}
    r(\mathbf{P}_{t},\mathbf{u}_{t})=\begin{cases}
H(\tilde{\mathbf{y}}_{k,t+1}\mid\tilde{\mathbf{y}}_{k,t}) & \mathbf{u}_{t}\in G(\mathbf{P}_{t})\\
-\infty & \text{o.w.}
\end{cases}
\end{equation}
where the negative infinity reward is assigned if an illegal action is taken at time $t$. To solve this MDP, we can apply the value iteration algorithm. Let $V_{\pi}(\mathbf{P}_{t})$  be the value function that defines the sum future reward when acting according to a certain policy $\pi$ starting from some state $\mathbf{P}_{t}$. Let $V^*(\mathbf{P}_{t})$ be the value function obtained using an optimal policy $\pi ^*$ that yields the maximum $V_{\pi}(\mathbf{P}_{t})$:  
\begin{equation}
    V^*(\mathbf{P}_{t}) = \max_{\mathbf{u}_{t}, \dots, \mathbf{u}_{T}} \sum_{t=1}^T r(\mathbf{P}_{t},\mathbf{u}_{t})
\end{equation}
We can express the optimal value using a recurrent relation as:
\begin{equation}
    \label{eq:value_iter}
    V^*(\mathbf{P}_{t}) =  \max_{\mathbf{u}_{t}} \left( r(\mathbf{P}_{t},\mathbf{u}_{t}) + V^*(T^{}(\mathbf{P}_{t}, \mathbf{u}_{t}))\right)
\end{equation}
Since $V^*(\mathbf{P}_{t})$ can be expressed using a recurrent relation, we can then use forward value iteration to solve for $V^*(\mathbf{P}_{1})$ and optimal ${\mathbf{u}_{1}, \dots, \mathbf{u}_{T}}$ \cite[p. 48]{lavalle2006planning}.   

\subsection{Computational complexity of path planning}
The main source of computational load in the proposed active sensing approach is the forward value iteration, which is used to solve for $V^*(\mathbf{P}_{1})$. Forward value iteration will have a complexity $O\left(TU^N \left( \frac{lw}{d^2} \right)^N \right)$, where $U^N$ is the size of the action space and $\left( \frac{lw}{d^2} \right)^N$ is the size of the state space.

\section{Results}
\label{sec:results}

\begin{figure}
    \centering  \includegraphics[width=0.9\linewidth]{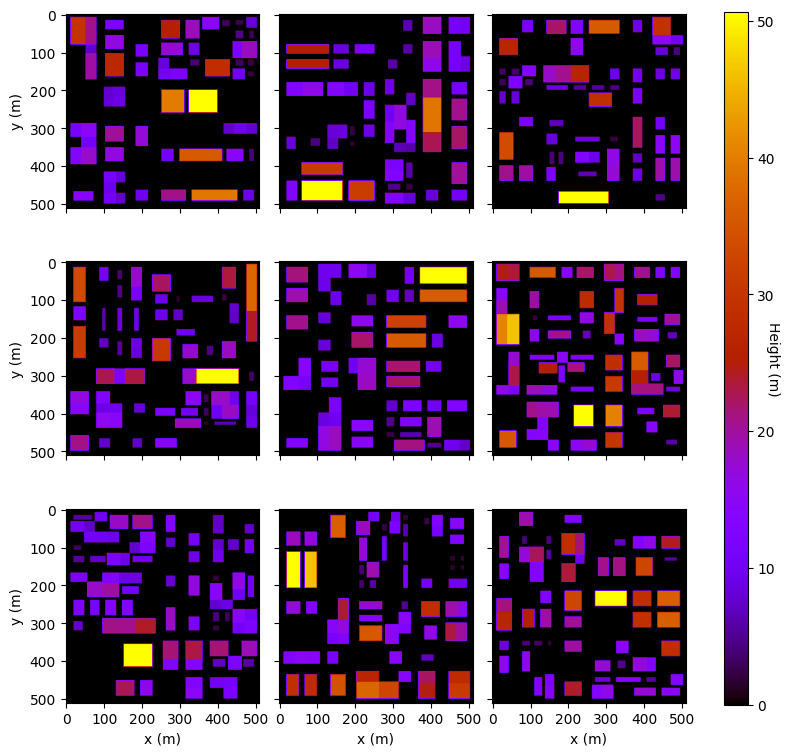}
    \caption{Heat maps of randomly generated urban environments. The colors corresponds to building heights. }
    \label{fig:example-enviornments}
\end{figure}

\subsection{Simulation environment}
\label{sec:dataset}
In this section, we describe the details of wireless channel simulation and the set of environments generated to train and test the proposed algorithms.

The main tool used for wireless channel simulation was ray tracing. Ray tracing is a channel propagation modeling tool that provides estimates of channel gain, angle of arrival/departure, and time delays by numerically solving Maxwell’s equations in far-field propagation conditions \cite{yun2015ray}. 
A ray-tracing software takes in the 3D map of the environment, along with other parameters, such as transmission frequency, transmitter location, and material properties of environment objects to trace the radio propagation paths and calculate the channel state at the desired points. The particular ray tracing software we used was Wireless Insite. 
To limit the ray-tracing computation time, we constrain the maximum number of reflections per a propagation ray to 3 and the maximum number of diffractions per ray to 1.

In order to create an expansive set of environments, we used a handcrafted script to randomly generate Manhattan-grid-like urban environments. 
We simulate a square shaped area of dimensions $486\text{m} \times 486$m. The generation procedure  starts by dividing the area into city blocks with random widths and length. 
The number of blocks per each dimension is 5, with a total of 25 blocks in the environment.
Then, open spaces and rectangular-base buildings with random dimensions are added within those blocks. 
Some examples of randomly generated environments are shown in Fig. \ref{fig:example-enviornments}.
In total, we generated 300 urban environments. In each environment, we placed transmitters uniformly spaced at 97.2m apart, which equates to 25 transmitter positions per environment. 
However, if a randomly generated transmitter location was indoor, it was removed from simulated environments. 
For each transmitter location, Wireless InSite was used to calculate the channel gain values over a 3D grid of points spaced at $4$ m apart and at altitudes ranging from 10 m to 30 m, over the entire width and length of the environment.
The calculations were ran for a carrier frequency of 5 GHz.
The dataset will be provided upon request to the authors.
The outputs from Wireless InSite were then processed in Python and used for simulations.
During training and testing, we crop the size of the simulated environment to a space with a footprint of size $384\text{m} \times 384$m with a random center within the original $486\text{m} \times 486$m area. This data augmentation was performed to add more diversity into the original data set and to add randomness to the transmitter locations.

\subsection{Training of DL channel gain predictor}

\begin{table}[t!]
    \caption{Architecture of the deep neural networks used for channel gain prediction.  }
    \label{tab:unet_table}
    \centering
    \resizebox{\linewidth}{!}{
    \begin{tabular}{|l|l|l|l|l|l|}
\hline \multicolumn{6}{|c|}{ \textbf{CG prediction U-Net }} \\
\hline \cellcolor{gray!25} \textbf{Layer} & \cellcolor{gray!25}\textbf{In} & $\cellcolor{gray!25}\mathbf{1}$ & $\cellcolor{gray!25}\mathbf{2}$ & $\cellcolor{gray!25}\mathbf{3}$ & $\cellcolor{gray!25}\mathbf{4}$ \\
\hline Out. size & $96 \times 96$ & $96 \times 96$ & $48 \times 48$ & $48 \times 48$ & $24 \times 24$ \\
\hline Channels & in & 16 & 16 & 32 & 32 \\
\hline Type & Conv. & Conv. & Conv. & Conv. & Conv. \\
\hline \cellcolor{gray!25} \textbf{Layer} & \cellcolor{gray!25}$\mathbf{5}$ & \cellcolor{gray!25}$\mathbf{6}$ & \cellcolor{gray!25}$\mathbf{7}$ & \cellcolor{gray!25}$\mathbf{8}$ &\cellcolor{gray!25} $\mathbf{9}$ \\
\hline Out. size & $24 \times 24$ & $12 \times 12$ & $12 \times 12$ & $6 \times 6$ & 4608 \\
\hline Channels & 64 & 64 & 128 & 128 & \\
\hline Type & Conv. & Conv. & Conv. & Conv. & Dense \\
\hline \cellcolor{gray!25}\textbf{Layer} &\cellcolor{gray!25} $\mathbf{1 0}$ & \cellcolor{gray!25}$\mathbf{1 1}$ & \cellcolor{gray!25}$\mathbf{1 2}$ & \cellcolor{gray!25}$\mathbf{1 3}$ & \cellcolor{gray!25}$\mathbf{1 4}$ \\
\hline Out. size & $12 \times 12$ & $24 \times 24$ & $48 \times 48$ & $96 \times 96$ & $96 \times 96$ \\
\hline Channels & 128 & 64 & 32 & 16 & 1 \\
\hline Skip connect. & 8 & 7 & 5 & 3 & 1 \\
\hline Type & Deconv. & Deconv. & Deconv. & Deconv. & Conv. \\
\hline
\end{tabular} }
\label{table:unet}
\end{table}
\begin{table}[t!]
    \centering
    \caption{Architecture of the deep neural networks used as \ac{DQN}s.}
    \label{tab:dqn}
\begin{tabular}{|l|l|l|l|l|}
\hline \multicolumn{5}{|c|}{ \textbf{DQN }} \\
\hline \cellcolor{gray!25}\textbf{Layer} & \cellcolor{gray!25}\textbf{In} & \cellcolor{gray!25}$\mathbf{1}$ & \cellcolor{gray!25}$\mathbf{2}$ & \cellcolor{gray!25}$\mathbf{3}$ \\
\hline Out. size & $96 \times 96$ & $48 \times 48$ & $24 \times 24$ & $16 \times 16$ \\
\hline Channel & in & 64 & 128 & 256 \\
\hline Filter size & & 4 & 4 & 2 \\
\hline Type & Conv. & Conv. & Conv. & Conv. \\
\hline \cellcolor{gray!25}\textbf{Layer} & \cellcolor{gray!25}$\mathbf{4}$ & \cellcolor{gray!25}$\mathbf{5}$ & \cellcolor{gray!25}$\mathbf{6}$ & \cellcolor{gray!25}$\mathbf{7}$ \\
\hline Out. size & 512 & 256 & 40 & 160 \\
\hline Type & Dense & Dense & Dense & Dense \\
\hline
\end{tabular}
\end{table}

\begin{table}[t!]
    \centering
    \caption{Training parameters for \ac{CG} predictor and DQN}
    \label{tab:training_parameters}
\begin{tabular}{|c|c|c|c|}
\hline \multicolumn{2}{|c|}{ \textbf{CG predictor training parameters} } & \multicolumn{2}{c|}{ \textbf{DQN training parameters }} \\
\hline \cellcolor{gray!25}\textbf{Description} & \cellcolor{gray!25}\textbf{Parameter} & \cellcolor{gray!25}\textbf{Description} & \cellcolor{gray!25}\textbf{Parameter} \\
\hline Learning rate & $10 ^{-3}$ & Learning rate & $10 ^{-5}$ \\
\hline Adam param. & $\beta_1=0.9$ & Exploration & $\varepsilon=0.03$ \\
\hline Adam param. & $\beta_2=0.999$ & Replay buffer & $6 \times 10^6$ \\
\hline Adam param. & $\hat{\epsilon}=10^{-6}$ & Batch size & 256\\
\hline Random walk param. & $p=0.8$ & Target update & $\tau_{DQN}=1000$ \\
\hline Batch size & $160$ & Learning steps & $M=5$ \\
\hline
\end{tabular}
\end{table}

Next, we describe the training details of the channel gain predictor proposed in Sec. \ref{sec:deep_learning_predictor}.
The data generated in 75 out of 300 city environments was used to train the predictor, while the data from 25 environments was used to validate the dataset. We refer to the former portion of the dataset as T1 and to the latter as T2. We used the Adam optimizer to minimize the loss function in Eq. \ref{eq:loss_func} \cite{kingma2014adam}. We found that training performance was highly dependent on the selection of the Adam parameters, which are shown in Table \ref{tab:training_parameters}, along with other relevant training parameters. We used the same notation for Adam parameters as in the original paper \cite{kingma2014adam}. In order to train the predictor, we randomly generated measurement inputs $\mathbf{y}_{k,{1:t}}$. 
During training, we assume measurements are obtained using a random waypoint motion model. We use a random trajectory to emulate measurement collection by UAVs on some planned paths. 
A random waypoint trajectory for a UAV is obtained as follows. 
At time $t$, each UAV takes independent random motion actions at probability $1-p$, and at probability $p$, the previous motion action is repeated by the UAV. 
The path length per UAV is random and uniformly distributed between 50 and 300 steps. 
Furthermore, we train separate \ac{DL} models depending on the number of UAVs $N$ collecting the measurements.

The architecture of the U-Nets used for ${\mu}_\theta(\mathbf{y}_{k,{1:t}}, \mathbf{M})$ and $  {\Sigma}_\theta(\mathbf{y}_{k,{1:t}}, \mathbf{M})$ is shown in Table \ref{table:unet}. The architecture consists of a series of convolutional layers or convolutional plus max-pooling layers to encode the inputs. 
The size of the output of each layer is shown in the table. 
Each layer outputs a number of channels which is equal to the number of convolutional filters in the layer. The convolutional filter size was $4 \times 4$, with stride size $1$.
A dense layer follows after the encoding layers. After, there is a sequence of layers that perform upsampling and convolution, which we refer to as deconvolution layers. The inputs of each deconvolution layer are concatenated with outputs of one of the encoding layers using skip connections. The skip connections are denoted in the table. 
The final layer is a convolutional layer which also uses skip connections.
We used the ReLU activation function for ${\mu}_\theta(\mathbf{y}_{k,{1:t}}, \mathbf{M})$ and tanh activation for $  {\Sigma}_\theta(\mathbf{y}_{k,{1:t}}, \mathbf{M})$. We found that using tanh activation for $  {\Sigma}_\theta(\mathbf{y}_{k,{1:t}}, \mathbf{M})$ leads to better performance than when using ReLU. 

\subsection{Training of DQN policies}
In this subsection, we describe the training details of the DQN policies for UAV control proposed in Sec. \ref{sec:deep_rl_controller}.
The data generated in 170 out of 300 city environments was used to train the algorithm while the data from 30 out of 300 environments was used to test the RL policy. We refer to the former portion of the dataset as T3 and to the latter as T4. 

We use the $\epsilon$-greedy policy for exploration, however the agent's random actions are steered. 
Namely, the agent never takes a random action that would lead to it leaving the map or colliding with a building. The value of $\epsilon$ is shown in Table \ref{tab:training_parameters}. We also ensure that the agent never leaves the map or collides with a building when taking actions according to the DQN or when moving randomly. 

The neural network architecture for \ac{DQN}s is shown in Table \ref{tab:dqn}. The DQNs consist of a series of convolutional layers with strides of size 4 or 2. The output size, the number of channels and filter size are shown in the Table \ref{tab:dqn}. The final layers of DQN are fully connected. The activation function used was ReLU.

\subsection{Benchmarks}
\label{sec:benchmarks}
Next, we explain the benchmark algorithms that we will compare our proposed approaches to. 
\subsubsection{Greedy active \ac{DL} prediction}The first benchmark is based on the \ac{CG} predictor that we introduced in Sec. \ref{sec:deep_learning_predictor}. The paths are designed to move the UAVs through the locations of maximum variance as predicted by  $\Sigma_\theta(\mathbf{y}_{k,{1:t}}, \mathbf{M})$. The intuition behind this approach is to collect new measurements in the locations where the predicted error is the largest. This also limits this approach to scenarios where $h_P=h_{\text{UAV}}$. Since the variance prediction $\Sigma_\theta(\mathbf{y}_{k,{1:t}}, \mathbf{M})$ is continuously updated by measurements obtained by the UAVs, the planned paths also need to be updated periodically over the course of time $1\leq t<T$. In the first $T_{start}=20$ steps, the UAVs move randomly since $\Sigma_\theta(\mathbf{y}_{k,{1:t}}, \mathbf{M})$ is unreliable for the purposes of path planning. Afterwards, UAV trajectories are updated every $T_{plan}=40$ steps. Let us denote the predicted covariance matrix for the set of locations $\mathcal{X}$ by $\left[ \Sigma_\theta(\mathbf{y}_{k,{1:t}}, \mathbf{M}) \right]_{\mathcal{X}}$. Then, the paths for the UAVs at time $t_1$ can calculated by solving the optimization problem:
\begin{subequations}
\begin{align} 
\label{eq:P4}
\tag{P4}
\max_{\mathbf{u}_{t_1}, \dots, \mathbf{u}_{T}} & \quad     \text{Tr} \left( \left[ \Sigma_\theta(\mathbf{y}_{k,{1:{T+1}}}, \mathbf{M}) \right]_{\mathcal{V}_{t_1:T+1}} \right) \\
\notag
\textrm{s.t.} & \quad 
\mathbf{P}_{t+1}=T(\mathbf{P}_{t}, \mathbf{u}_{t}), \mathbf{u}_t \in G(\mathbf{P}_t)
\end{align}
\end{subequations}
As with problem (\ref{eq:P3}), we convert this problem into an MDP. In order to maximize the objective function in (\ref{eq:P4}), it is necessary to keep track of the locations visited by the UAVs to avoid repeated visits. This can be achieved by defining the state $\mathbf{s}_{t}$ to include all locations visited up to time $t$. However, in this case, the size of the state space would be too large for efficient computation of optimal paths. Instead, we ensure that UAVs do not perform repeated visits by appropriately designing the reward function. The reward function at time $t$ is defined as:
\begin{multline}
    r(\mathbf{P}_{t},\mathbf{u}_{t})=\\
\begin{cases}
\text{Tr} \left( \left[ \Sigma_\theta(\mathbf{y}_{k,{1:{t+1}}}, \mathbf{M}) \right]_{\mathcal{V}_{t:t+1}} \right) & \mathbf{u}_{t}\in G^*(\mathbf{P}_{t}, \mathbf{P}_{t_1})\\
-\infty & \text{o.w.}
\end{cases}
\end{multline}
where the function $G^*(\mathbf{P}_{t}, \mathbf{P}_{t_1})$ ensures that no illegal actions are taken and also that the UAVs are moving away from their respective starting locations. The latter is necessary to ensure that UAVs are not visiting the same location multiple times. The state at time $t$ is defined as  $\mathbf{s}_{t}=\mathbf{P}_{t}$ and the action is $\mathbf{a}_{t}=\mathbf{u}_{t}$. Given this MDP definition, we can calculate the UAV paths using value iteration (Eq. \ref{eq:value_iter}). Furthermore, due to the nature of the reward function, forward value iteration can be applied independently per UAV.

We will use this benchmark to compare against our proposed active \ac{DL} \ac{CG} prediction approach, since it is also transmitter location free and uses 3D maps.
The disadvantage of this benchmark compared to the proposed active \ac{DL} approach is that the greedy objective function in (\ref{eq:P4}) can lead to multiple UAVs exploring locations in close proximity of one another if these locations have high variance as predicted by $\Sigma_\theta(\mathbf{y}_{k,{1:t}}, \mathbf{M})$. Furthermore, as discussed in Sec. \ref{sec:deep_rl_controller}, since DNNs are black-box models, collecting the measurements in regions of highest predicted variance may not minimize the final prediction error.

The main source of computational load in this benchmark is the forward value iteration, which has a computational complexity $O\left(NTU \left( \frac{lw}{d^2} \right) \right)$.

\subsubsection{Greedy active Kriging prediction} The second benchmark is based on the Kriging predictor explained in Sec. \ref{sec:kriging_interpolation}. Similar to our previous benchmark, the paths are designed to move the UAVs through the locations of maximum variance as predicted by  $\tilde{\mathbf{\Sigma}}_{k, 1:t}$ in Eq. \ref{eq:kriging_variance}. 
\begin{subequations}
\begin{align} 
\label{eq:P5}
\tag{P5}
\max_{\mathbf{u}_{t_1}, \dots, \mathbf{u}_{T}} & \quad     \text{Tr} \left( \left[ \tilde{\mathbf{\Sigma}}_{k, 1:T+1}\right]_{\mathcal{V}_{t_1:T+1}} \right) \\
\notag
\textrm{s.t.} & \quad 
\mathbf{P}_{t+1}=T(\mathbf{P}_{t}, \mathbf{u}_{t}), \mathbf{u}_t \in G(\mathbf{P}_t)
\end{align}
\end{subequations}
The path calculation is performed using forward value iteration in the same way as in the previous benchmark. 
We will use this benchmark to compare against our proposed active Kriging \ac{CG} prediction approach, since it also requires transmitter location to be known.
This benchmark has the disadvantage that it can lead to multiple UAVs exploring similar regions due to the greedy objective function in (\ref{eq:P5}).
\subsubsection{Random waypoints with DL prediction approach}
In this benchmark approach, UAVs move according to the random waypoints strategy and prediction is done using our proposed DL predictor, as explained in  Sec. \ref{sec:deep_learning_predictor}. The purpose of this approach is to evaluate the importance of optimal path planning for \ac{CG} prediction. We will use this benchmark to quantitatively compare against our proposed active \ac{DL} \ac{CG} prediction approach. Furthermore, we will use it to evaluate the accuracy of the \ac{CG} predictor for various scenarios in the absence of optimized path planning.

\subsubsection{Random waypoints with Kriging prediction approach}
In this benchmark approach, UAVs move according to the random waypoints strategy and prediction is done using Kriging prediction, explained in Sec. \ref{sec:kriging_interpolation}. We will use this approach as a benchmark to compare against our proposed active Kriging prediction approach. Furthermore, we will use it to evaluate the accuracy of Kriging prediction for various scenarios in the absence of optimized path planning.

\subsection{Evaluation of the \ac{DL} \ac{CG} predictor}

\begin{figure}
    \centering
    \includegraphics[width=0.9\linewidth]{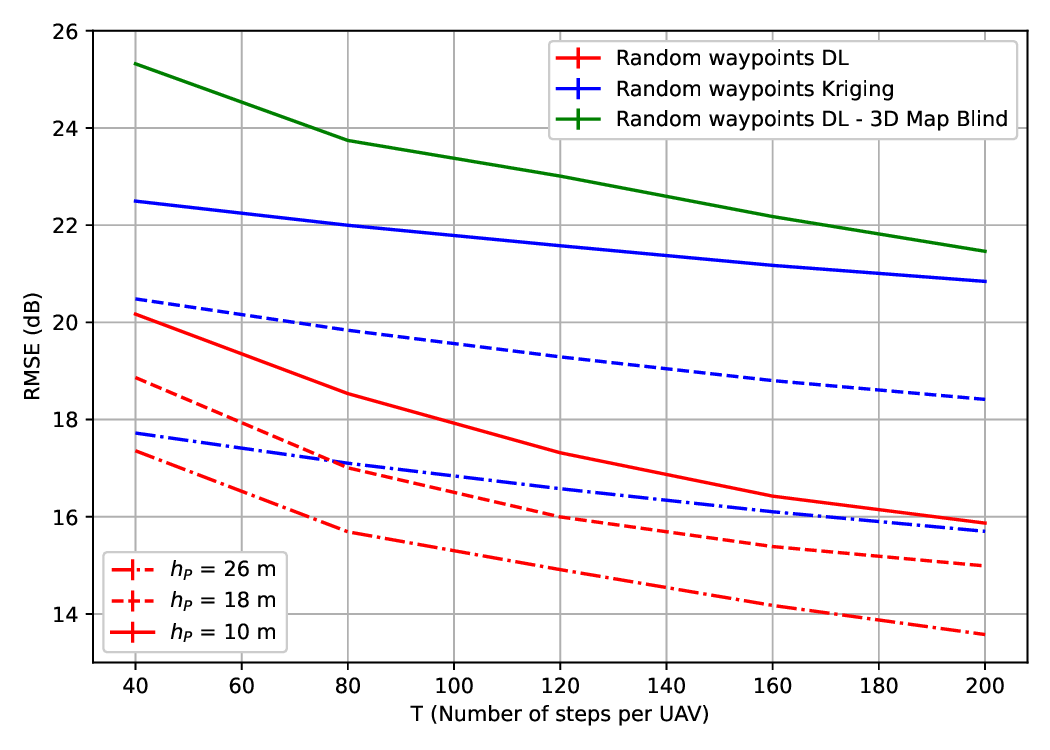}
    \caption{RMSE of \ac{CG} prediction for different prediction altitudes $h_P$ and different number of moved steps per UAV $T$. 
    The measurements are collected on random UAV paths using 3 UAVs.}
    \label{fig:error-per-height}
\end{figure}

First, we evaluate the performance of the probabilistic DL \ac{CG} predictor without optimized path planning and use random-waypoints UAV motion with $p=0.8$ for measurement collection. There are $N=3$ UAVs collecting the measurements. The starting location of the UAVs is randomized within a randomly placed $40$m $\times 40$m rectangle. This simulates a UAV swarm being deployed from a common starting area. We use the RMSE as the metric to evaluate the accuracy of \ac{CG} prediction. We only evaluate the accuracy at unvisited locations $\mathcal{Q}_P \backslash \mathcal{V}_{1:T}$, since the accuracy at visited locations is perfect due to the assumption of noiseless measurements. We define a utility binary vector variable $\tilde{\mathbf{z}} \in \mathbb{Z}_2^{|\mathcal{Q}_P|}$, where $[\tilde{\mathbf{z}}]_j=0$ if the location $\tilde{\mathbf{q}}_j \in \mathcal{Q}_P$ is obstructed by a building or if it is in $\mathcal{V}_{1:T}$, and $[\tilde{\mathbf{z}}]_j=1$ otherwise. Then, the RMSE is defined as: 
$\sqrt{
\frac{1}{||\tilde{\mathbf{z}}||_1}\Delta_k^T \text{diag}\left(\tilde{\mathbf{z}}\right)
\Delta_k}
$. We show the RMSE as a function of number of steps $T$ per UAV in Fig. \ref{fig:error-per-height} on T4 dataset. We compare our \ac{DL} \ac{CG} predictor against Kriging interpolation and a 3D-map-blind predictor. The 3D-map-blind approach is identical to our proposed DL approach except it does not use 3D maps as an input. We evaluate the prediction methods for different prediction altitudes $h_P$. First, we can observe that the 3D-map-blind approach performs significantly worse compared to the proposed approach for $h_P= 10$ m, which is why we do not evaluate it for other altitudes. This implies that our proposed approach successfully uses 3D maps for prediction and also that 3D maps are particularly useful when transmitter location is unknown. Furthermore, the proposed \ac{DL} \ac{CG} approach performs significantly better than Kriging interpolation, even though Kriging interpolation relies on transmitter location. This is achieved through the use of 3D maps and deep learning for \ac{CG} prediction. The gap between Kriging interpolation and the proposed \ac{DL} \ac{CG} predictor decreases with increasing $h_P$. This is likely due to the fact that at higher altitudes, the channel gain is easier to predict due to line-of-sight channel being more common between receiver and transmitter.

\begin{figure}
    \centering
    \includegraphics[width=0.9\linewidth]{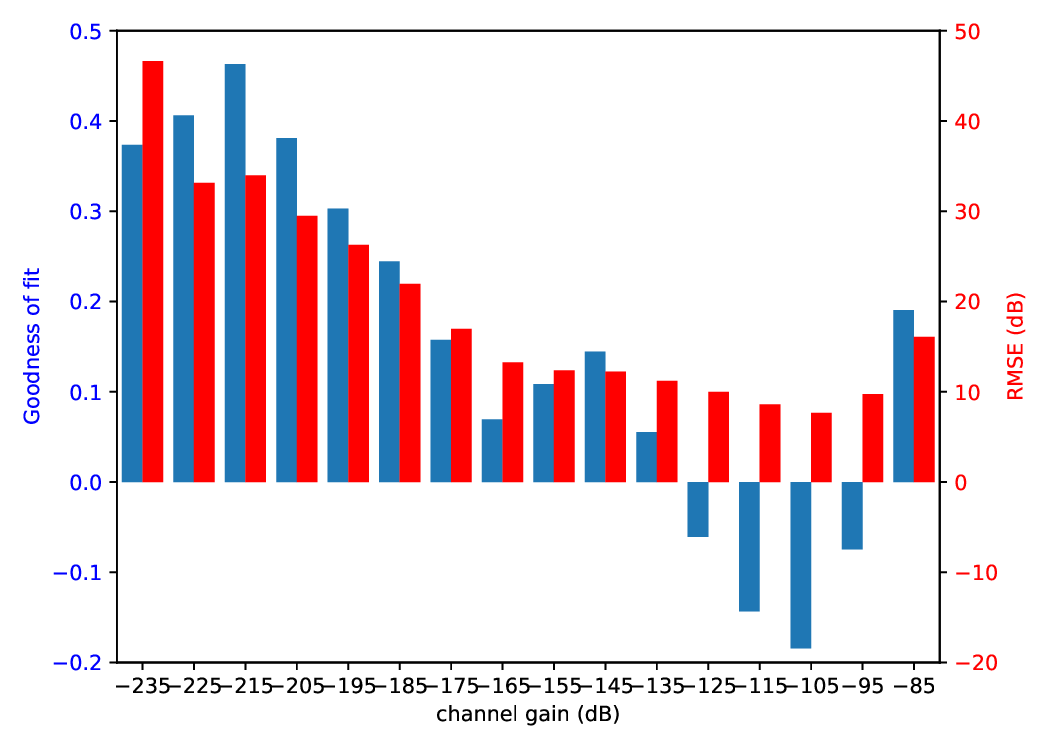}
    \caption{RMSE and goodness of fit of \ac{DL} \ac{CG} predictor for different \ac{CG} values.}
    \label{fig:goodness-of-fit}
\end{figure}

Next, we evaluate the accuracy of the \ac{DL} \ac{CG} predictor as a function of \ac{CG} after $T=200$ steps per UAV. In Fig. \ref{fig:goodness-of-fit}, red bars correspond to the RMSE for different \ac{CG} value bins. There are 16 \ac{CG} bins in the figure between -240 dB and -80 dB. The figure is obtained by grouping all of the locations in T4 environments based on their corresponding \ac{CG} bin and then taking the RMSE in each group. From the figure, we observe that the prediction RMSE is lower for higher \ac{CG} values. 
This likely occurs because there are more training points for higher \ac{CG} values in the training data, which skews the accuracy of the predictor towards higher \ac{CG} values. We also evaluate the accuracy of variance prediction $\Sigma_\theta(\mathbf{y}_{k,{1:T}}, \mathbf{M})$. We introduce a goodness of fit metric which is equal to the log of average ratio of the square prediction error over the predicted variance at any location $j$:
$
\log \mathbf{E}_j \left[ \left( {[\Delta_k]_j ^2} \right)/ \left({[\Sigma_\theta(\mathbf{y}_{k,{1:t}}, \mathbf{M})]_{j,j}} \right) \right]
$.
The goodness of fit value should be ideally close to 0, which would happen if variance prediction is equal to the observed error. Low absolute value of goodness of fit is necessary for the output $\Sigma_\theta(\mathbf{y}_{k,{1:T}}, \mathbf{M})$ to be useful for the proposed path planning algorithms. The blue bars in Fig. \ref{fig:goodness-of-fit} correspond to the goodness of fit values for different \ac{CG} bins. The absolute values of goodness of fit are close to 0 across all \ac{CG} bins and are generally positive, which indicates that the predicted variance is on average lower than the actual error. Overall, the absolute value of goodness of fit is lower for lower \ac{CG} values, where the RMSE is also high. 

\subsection{Computational delay of proposed active \ac{CG} prediction approaches}

\begin{figure}
    \centering
    \includegraphics[width=0.9\linewidth]{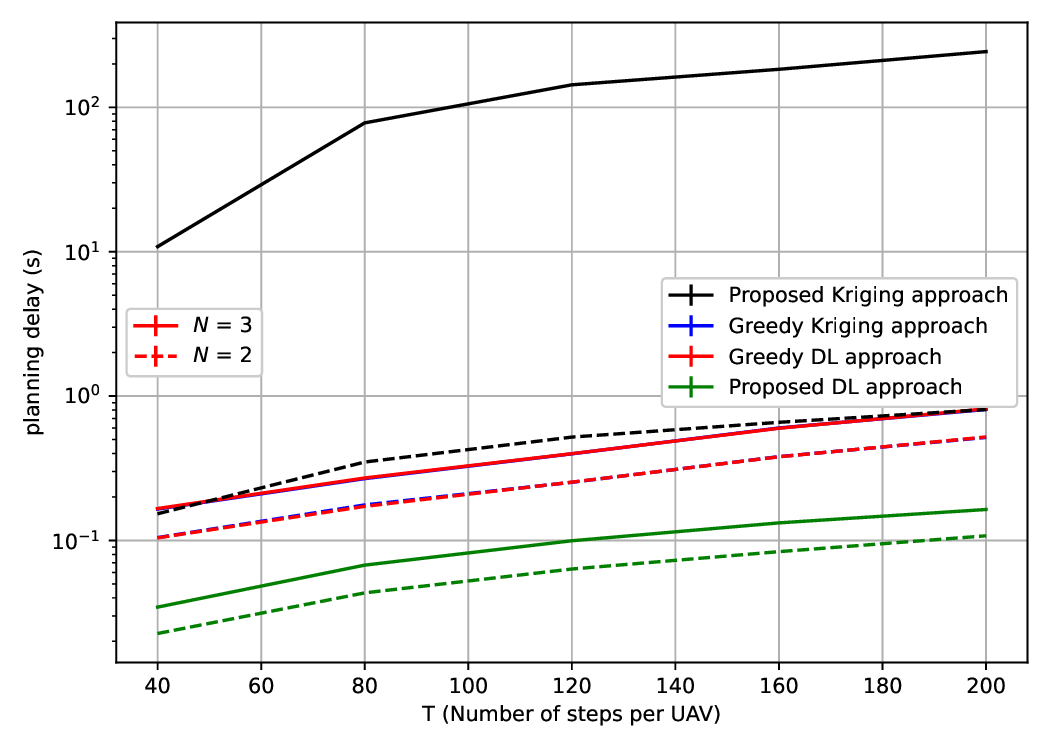}
    \caption{Compute delay of path planning for proposed active prediction approaches for different numbers of UAVs $N$.}
    \label{fig:planning_delay}
\end{figure}
We evaluate the proposed active prediction approaches in terms of path planning compute delay measured in seconds for $N=2$ and $N=3$ (Fig. \ref{fig:planning_delay}). The results are obtained on a workstation with a AMD Ryzen Threadripper PRO 5975WX 32-Core CPU and an NVIDIA GeForce RTX 3090 GPU. The GPUs were used for execution of neural networks whenever applicable. The active prediction approaches were implemented in Python. Path planning was accelerated by transforming various path-planning  operations such as forward value iteration into array or matrix operations using NumPy library. Tensorflow library was used for implementation of \ac{DL} components. The greedy Kriging and \ac{DL} approach have identical computational delay since the path planning algorithms have identical computational complexity. 
The path planning delay for the proposed \ac{DL} approach is due to the delay of the RL policy {DNN}, so it is dependent on the size of the DNN and the GPU used for execution. On our workstation, the proposed \ac{DL} approach is significantly faster than the greedy approaches. The highest delay approach is the proposed Kriging approach, whose complexity scales exponentially with $N$. Given our Python implementation and capabilities of our workstation, running the proposed Kriging approach for $N>3$ is not feasible. The complexity of path-planning of this  approach could be reduced by down-sampling the AoI $\mathcal{Q}$ to reduce the state space size or by dividing the UAV swarm into clusters of UAVs whose path planning is performed independently. However, this is beyond the scope of this paper and so we limit the evaluation of the proposed Kriging approach to $N \leq 3$.

\subsection{Evaluation of proposed active \ac{CG} prediction approaches}

\begin{figure*}
  \centering
  \begin{minipage}{.32\linewidth}
    \centering
    \includegraphics[width=\linewidth]{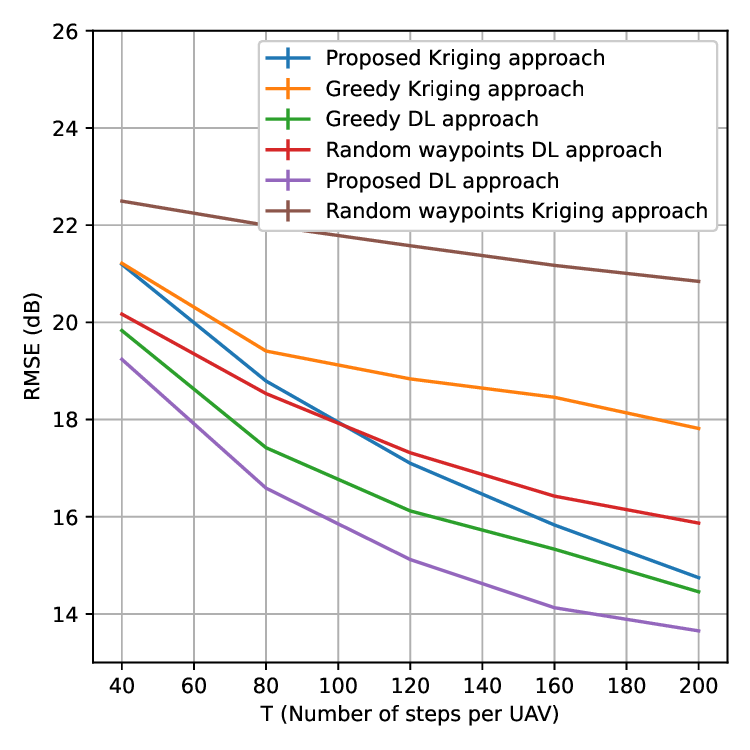}
    \caption{Prediction RMSE for three UAVs for the proposed approaches and the benchmarks. The starting location of the UAVs is randomized within a randomly placed $40$m $\times 40$m rectangle.}
    \label{fig:error-3uav}
  \end{minipage}\quad
  \begin{minipage}{.32\linewidth}
    \centering
    \includegraphics[width=\linewidth]{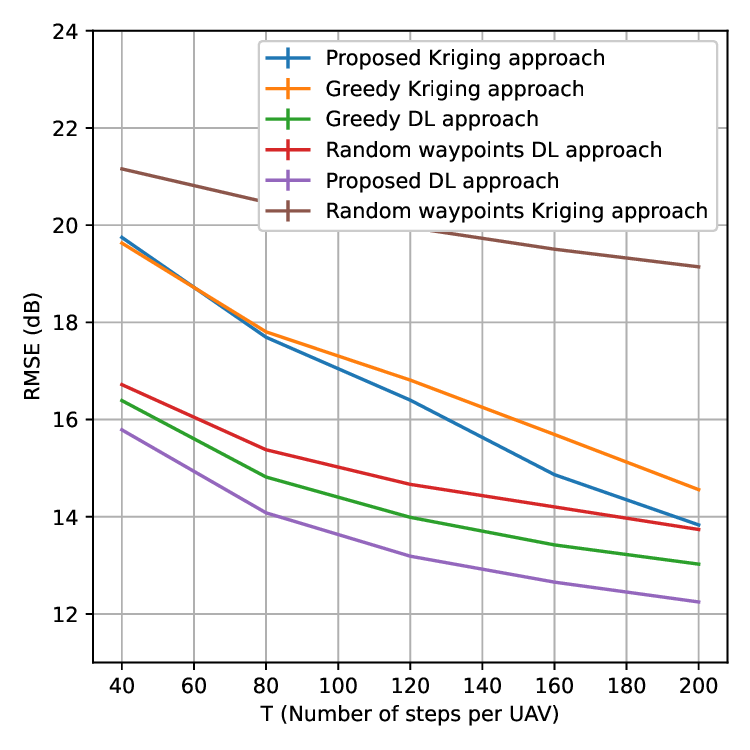}
    \caption{Prediction RMSE for three UAVs for the proposed approaches and the benchmarks. The starting location of the UAVs is randomized within the entire AoI.}
    \label{fig:error-3uav-sep}
  \end{minipage}\quad
  \begin{minipage}{.32\linewidth}
    \centering
    \includegraphics[width=\linewidth]{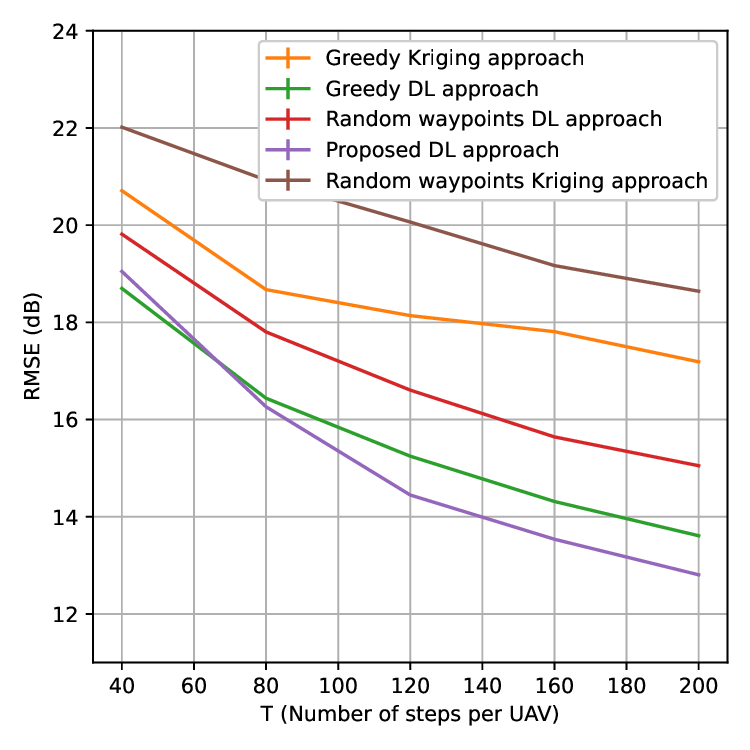}
    \caption{Prediction RMSE for five UAVs for the proposed approaches and the benchmarks. The starting location of the UAVs is randomized within a randomly placed $40$m $\times 40$m rectangle.}
    \label{fig:error-5uav}
  \end{minipage}\quad
\end{figure*}

Next, we evaluate the performance of the proposed active prediction approaches in terms of the prediction RMSE and compare them to the benchmarks described in Sec. \ref {sec:benchmarks}. In Fig. \ref{fig:error-3uav}, we display the results for three coordinated UAVs for $h_P=10$m. The starting location of the UAVs is randomized within a randomly placed $40$m $\times 40$m rectangle. The proposed active \ac{DL} \ac{CG} prediction approach and the proposed active Kriging approach outperform their greedy and random waypoints benchmarks. We can observe a significant gap in RMSE between proposed approaches and their random waypoints benchmarks, which demonstrates the importance of optimal path planning for measurement collection. The proposed active Kriging approach also outperforms the greedy Kriging benchmark. The gap exists because the proposed active Kriging approach design paths that maximize the joint entropy of measured \ac{CG}s instead of independently moving the UAVs towards locations with highest predicted variance. For similar reasons, the proposed active \ac{DL} prediction approach that relies on \ac{RL} for path planning outperforms the greedy \ac{DL} benchmark. Furthermore, greedy measurement collection may not be optimal for \ac{DL}-based predictors since we do not know how a deep neural network predicts channel gain, therefore \ac{RL}-based measurement collection can have an advantage over greedy measurement collection. Overall, the proposed active \ac{DL} prediction method performs better than the proposed active Kriging prediction method in terms of RMSE. However, both proposed methods have practical advantages. Kriging-based active prediction has the advantage of not requiring extensive training data and 3D map knowledge, while the proposed DL approach does not require the knowledge of transmitter location and provides higher accuracy.

We also evaluate the proposed algorithms and the benchmarks for scenarios when the starting locations of the UAVs are randomized across the entire AoI. This can for example simulate the case when UAVs have been previously deployed to complete different tasks and have moved far apart before commencing collection of \ac{CG} measurements. These results are shown in Fig. \ref{fig:error-3uav-sep}. We can see that the RMSE across all approaches decreases, which occurs because the UAVs are more spread out across the AoI. Moreover, the gap between the proposed approaches and greedy benchmarks decreases due to UAVs being more likely to move in non-overlapping areas since their starting locations are far apart. 
In Fig. \ref{fig:error-5uav}, we show the results for 5 UAVs with starting locations randomized within a randomly placed $40$m $\times 40$m rectangle. The RMSE across all approaches decreases compared to the results with 3 UAVs. Moreover, the gap between the proposed active \ac{DL} prediction approach and its greedy benchmark decreases compared to the scenario with 3 UAVs. This indicates that for \ac{DL} prediction, coordination is less important for a larger number of UAVs.

\section{Conclusions}
\label{sec:conclusions}

In this paper, we developed methods for prediction of \ac{CG} that use environment-specific features such as building maps and \ac{CG} measurements to achieve a high level of prediction accuracy.
We assume that measurements are collected using a swarm of coordinated UAVs.
We developed two active prediction approaches based on \ac{DL} and Kriging interpolation. 
We trained and evaluated the two proposed approaches in a ray-tracing-based channel gain simulator. 
Using channel simulations based on the ray-tracing approach, we demonstrated the importance of active prediction compared to prediction based on randomly collected measurements of channel gain. Furthermore, we showed that using \ac{DL} and 3D maps, we can achieve high prediction accuracy even without knowing the transmitter location. We also demonstrated the importance of coordinated path planning for active prediction when using multiples UAVs compared to UAVs collecting measurements independently in a greedy manner.

\bibliography{references}
\bibliographystyle{ieeetr}

\end{document}